\documentclass[twocolumn]{article}

\usepackage[utf8]{inputenc} 
\usepackage[T1]{fontenc}    
\usepackage[margin=0.75in]{geometry}
\usepackage{microtype}
\usepackage{graphicx}
\usepackage{booktabs}       
\usepackage{amsfonts}       
\usepackage{nicefrac}       
\usepackage{microtype}      
\usepackage{times}
\usepackage{tikz,pgf}
\usepackage{epsfig,amssymb,amsmath,ifthen,comment}
\usepackage{dsfont}
\usepackage{xr}
\usepackage{hyperref}
\usepackage{mathtools}

\renewcommand{\cite}{\citep}

\newenvironment{prf}{\noindent\textit{Proof:}\begin{mdseries}}{\end{mdseries}{\hfill\scriptsize$\Box$}} 

\newcommand\numberthis{\addtocounter{equation}{1}\tag{\theequation}}
\newcommand{\G}{{\cal G}}

\newcommand\mscriptsize[1]{\mbox{\scriptsize\ensuremath{#1}}}

\newtheorem{lem}{Lemma}

\newtheorem{assump}{Assumption}

\DeclareMathOperator{\doo}{do}
\DeclareMathOperator{\dis}{dis}

\DeclareMathOperator{\pa}{pa}
\DeclareMathOperator{\de}{de}
\DeclareMathOperator{\si}{si}
\DeclareMathOperator{\ch}{ch}
\DeclareMathOperator{\an}{an}

\DeclareMathOperator{\mb}{mb}

\DeclareMathOperator{\pre}{pre}

\DeclareMathOperator{\pas}{pa^s}

\newenvironment{lema}[1]{\par\noindent{\bf Lemma #1\ }\em}{\em}

\usepackage[round]{natbib}

\title{Path Dependent Structural Equation Models}
\author{Ranjani Srinivasan\textsuperscript{1}, Jaron Lee\textsuperscript{2}, Rohit Bhattacharya\textsuperscript{2}, Narges Ahmidi\textsuperscript{2,3}, Ilya Shpitser\textsuperscript{2}\\
	{\normalsize \textsuperscript{1}Electrical and Computer Engineering, Johns Hopkins University} \\
	{\normalsize \textsuperscript{2}Computer Science, Johns Hopkins University}\\
	{\normalsize \textsuperscript{3}Institute of Computational Biology, Helmholtz Munich Center}\\
	{\normalsize Corresponding author: Ranjani Srinivasan (ranjani@jhu.edu)}
		}

\begin{document}
\maketitle

\begin{abstract}
 Causal analyses of longitudinal data generally assume that the qualitative causal structure relating variables remains invariant over time.
 In structured systems that transition between qualitatively different states in discrete time steps, such an approach is deficient on two fronts. First,  time-varying variables may have
 \emph{state-specific} causal relationships that need to be captured. Second, an intervention can result in state transitions downstream of the intervention
 different from those actually observed in the data.  In other words, interventions may counterfactually alter the subsequent temporal evolution of the system. We introduce a generalization of causal graphical models, Path Dependent Structural Equation Models (PDSEMs), that can describe such systems.
 We show how causal inference may be performed in such models and illustrate its use in simulations and data obtained from a septoplasty surgical procedure. 

\end{abstract}

\section{Introduction}
Many scientific questions and engineering tasks may only be approached by analyzing the behavior of a system over time. 
Understanding long term health risk factors \citep{belangerNursesHealthStudy1978}, trajectory tracking \citep{richards2005fundamentals}, speech recognition \citep{rabiner1989tutorial}, game playing \citep{thrun1995learning}, all require modeling the temporal evolution of a system. Many models for longitudinal or time series data, such as hidden Markov models or Kalman filters, are graphical models, and most may be viewed as dynamic Bayesian networks (DBNs) \citep{murphyMachineLearningProbabilistic2012}.
These models are used to predict the future evolution of systems, 
or find latent structures that best explain observations. 
Despite their complexity and usefulness,
these models deal with fundamentally \textit{associative} relationships. However,
testing empirical hypotheses
or providing decision support tools in complex
domains that vary over time often requires \textit{causal} modeling. 

Causal models previously used for such tasks include graphical causal models \citep{pearl09causality},
marginal structural models \citep{robins97marginal}, structural nested models \citep{robins99marginal},
as well as models of counterfactual regret \citep{murphy2003optimal,Moodie13DTR}.
However, these models have
generally assumed an invariant 
causal structure 
over time.
For example, analysis of the impact of anti-retroviral therapy
on HIV infection progression in observational studies assumed the same variables relevant for the
patient health and the same causal relationships linking them at each time point in the study \citep{hernan2000marginal}.
Changes tracked over time (such as HIV developing resistance to the current drug) are thus \textit{quantitative},
with the underlying causal structure remaining unchanged over time. However, 
many systems undergo \textit{qualitative} changes as well, where observability, relevance, and
causal relationships of variables vary over time.

Consider the task of modeling surgical procedures to make informed decisions on resident surgeon training.
Surgeries are often divided into discrete stages, each with an intermediate goal \citep{ahmidi2015automated}.
Each stage is associated with a distinct set of 
variables and relationships among them that may not be shared across stages.
For instance, stitching together a previously made incision 
is a routine task requiring few tools that 
may be
executed by a surgical robot, while reconstructing cartilage 
may require multiple tools, high surgical skill and manual dexterity.
Another feature of surgeries is that procedures performed at a particular stage can 
go wrong, forcing surgeons to "double back" to 
correct mistakes, or deal with complications.  
Surgeon experience may often determine whether 
previous stages of the surgery are revisited. 
The goal of causal inference in this setting is to help assign surgeons to perform different stages of the surgery while navigating the tradeoff between the need to train resident surgeons on the one hand, and operating costs and patient safety on the other.

Addressing this tradeoff entails using retrospective data to estimate outcomes of surgery trajectories that \emph{differ from those actually observed} due to counterfactually different choices of surgeon assignment in past stages of the surgery. Following the convention in the economics literature, we call the phenomenon where the evolution of a system changes in response to counterfactually different past choices
\textit{path dependence} \citep{liebowitzPathDependence2002}. 

Other examples of path dependence include life course studies examining economic disparities in society or patient outcomes in hospitals using Electronic Health Record (EHR) data. Whether and where subjects went to elementary and high school, college, and their place of work, all represent qualitatively different stages of subjects' lives, with correspondingly different variables and causal relationships among them.  In addition, subjects might re-enroll in school, or otherwise revisit different stages in their life.  Finally, life outcomes and life trajectories depend on counterfactually different life choices in the past, such as the choice to not enroll in college.
Similar arguments can be made for patients being assigned to different parts of a hospital, such as the emergency room, the ward, the intensive care unit, or the operating room.  Causally relevant variables and relationships between them differ drastically depending on which part of the hospital patients find themselves in.  In practice, patients may revisit above parts of the hospital multiple times, often in response to various treatment decisions made.  Counterfactually different decisions would result in potentially different patient trajectories.

In this paper, we introduce the \emph{path-dependent structural equation model (PDSEM)} for causal systems that exhibit qualitative changes over time, observed or unobserved confounding, \emph{and} path-dependence on counterfactual choices in the past.
Our model can be viewed as a generalization of causal dynamic Bayesian networks, that allows complex and repeating stage transitions, and distinct causal models at each stage,
{or as a generalization of Markov decision processes (MDPs) that uses causal models in each state.  In our formulation, each state is modeled explicitly as a (possibly hidden variable) causal model, and where observed actions are not chosen by the optimizer as in reinforcement learning problems, but instead determined by the data generating process that may contain unobserved confounders.  As a result, our model combines complex and potentially looping state transitions of MDPs, and complex relationships among variables 
	of a causal model.  {PDSEMs may also be viewed as a generalization of a Markov chain endowed with graphical causal model semantics, which allows handling of confounding and analysis of counterfactual state transitions.}

\section{Background} \label{bkgd}

We first introduce necessary causal modeling ideas and previous work, before extending them to allow path-dependence.

\subsection{Statistical and Causal Graphical Models} \label{sec:dag-model}

The statistical model of a directed acyclic graph (DAG) ${\cal G}({\bf V})$ with a vertex set ${\bf V} \equiv \{ V_1, \ldots, V_k \}$, called a \textit{Bayesian network}, is the set of distributions that 
factorize with respect to the DAG as $p({\bf V}) = \prod_{V_i \in {\bf V}} p(V_i \mid \pa_{\cal G}(V_i))$ where $\pa_{\cal G}(V_i)$ are parents of $V_i$ in ${\cal G}$.

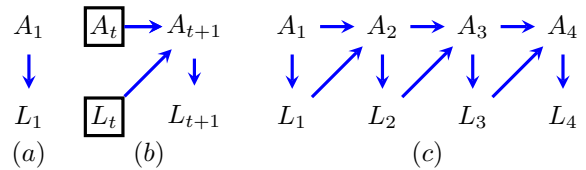
\begin{figure}  %
	\begin{center}
		\begin{tikzpicture}[>=stealth, node distance=1.2cm]
		\tikzstyle{format} = [draw=none, very thick, circle, minimum size=7.0mm,
		inner sep=0pt]
		\tikzstyle{square} = [draw, very thick, rectangle, minimum size=5.0mm, inner sep=0pt]
			
		\begin{scope}[xshift=-1.8cm,yshift=-2.5cm]
		\path[->, very thick]
		node[format] (a) {$A_1$}
		node[format, below of=a] (l) {$L_1$}
		
		(a) edge[blue] (l)
		
		node[below of=l, yshift=0.7cm, xshift=0.0cm] (l) {$(a)$}
		;
		\end{scope}
		
		\begin{scope}[xshift=-.8cm,yshift=-2.5cm]
		\path[->, very thick]
		node[square] (a) {$A_t$}
		node[square, below of=a] (l) {$L_t$}
		
		node[format, right of=a] (a2) {$A_{t+1}$}
		node[format, below of=a2] (l2) {$L_{t+1}$}
		
		(l) edge[blue] (a2)
		(a) edge[blue] (a2)
		(a2) edge[blue] (l2)
		(a) edge[blue] (a2)
		
		node[below of=l, yshift=0.7cm, xshift=.6cm] (l) {$(b)$}
		;
		\end{scope}
		
		\begin{scope}[xshift=1.7cm,yshift=-2.5cm]
		\path[->, very thick]
		node[format] (a0) {$A_1$}
		node[format, below of=a0] (l0) {$L_1$}
		
		node[format, right of=a0] (a1) {$A_{2}$}
		node[format, below of=a1] (l1) {$L_{2}$}
		
		node[format, right of=a1] (a2) {$A_{3}$}
		node[format, below of=a2] (l2) {$L_{3}$}
		
		node[format, right of=a2] (a3) {$A_{4}$}
		node[format, below of=a3] (l3) {$L_{4}$}
		(a0) edge[blue] (a1)
		(a1) edge[blue] (a2)
		(a0) edge[blue] (l0)
		(a1) edge[blue] (l1)
		(a2) edge[blue] (l2)
		(a3) edge[blue] (l3)
		(a2) edge[blue] (a3)
		
		(l0) edge[blue] (a1)
		(l1) edge[blue] (a2)
		(l2) edge[blue] (a3)
		
		node[below of=l2, yshift=0.7cm, xshift=-.6cm] (l) {$(c)$}
		;
		\end{scope}

	\end{tikzpicture}
	\end{center}
	\caption{
	(a) Prior network DAG $\G_0$, representing the state of the dynamic Bayesian network at time $t=0$.
	(b) A conditional DAG $\G_{t,t+1}$ representing the transitions in a dynamic Bayesian network.
	(c) A dynamic Bayesian network model unrolled to four time steps.
	}
	\label{fig:dynamic}
\end{figure}

Causal models of a DAG are also sets of distributions but on \textit{counterfactual} random variables. 
Each variable $V_i$ in a causal model is determined from values of its parents $\pa_{\cal G}(V_i)$ and an exogenous noise variable $\epsilon_i$ via an invariant causal mechanism called a \emph{structural equation} $f_i(\pa_{\cal G}(V_i), \epsilon_i)$.  Causal models allow counterfactual intervention operations, denoted by the $\text{do}({\bf a})$ operator in
\citep{pearl09causality}.  Such operations replace each structural equation $f_i(\pa_{\cal G}(V_i), \epsilon_i)$ for $V_i \in {\bf A} \subset {\bf V}$ by one that sets $V_i$ to a constant value in ${\bf a}$ corresponding to $V_i$.  The joint distribution of variables in ${\bf Y} \equiv {\bf V} \setminus {\bf A}$ after the intervention $\text{do}({\bf a})$ was performed is denoted by
$p({\bf Y} \mid \text{do}({\bf a}))$, equivalently written as $p(\{ V_i({\bf a}) : V_i \in {\bf Y} \})$, or $p({\bf Y}({\bf a}))$, where $V_i({\bf a})$
is a counterfactual random variable or a potential outcome.

A popular causal model called the \emph{non-parametric structural equation model with independent errors (NPSEM-IE)} \citep{pearl09causality} assumes, 
aside from the structural equations for each variable
being functions of their parents in the DAG ${\cal G}({\bf V})$, that 
the joint distribution of all exogenous terms are marginally independent: $p(\epsilon_1, \epsilon_2, \ldots) = \prod_{V_i \in {\bf V}} p(\epsilon_i)$.  The NPSEM-IE implies the DAG factorization of $p({\bf V})$ with respect to ${\cal G}({\bf V})$, and a truncated DAG factorization known as the \emph{g-formula}:

\vspace{-0.55cm}
{\small 
\begin{align}
p({\bf Y}({\bf a})) =
\prod_{V_i \in {\bf Y}} p(V_i \mid \pa_{\cal G}(V_i))
\vert_{ {\bf A} = {\bf a} }
\label{eqn:g}
\end{align}
}
for every ${\bf A} \subseteq {\bf V}$, and ${\bf Y} = {\bf V} \setminus {\bf A}$.

\subsection{{Graphical Models In Discrete Time}}
\label{sec:dbn-def}

While Bayesian networks 
lend themselves well to the modeling of static data, 
dynamic data with temporal evolution 
requires more sophisticated 
models.
A generalization of the Bayesian network model for discrete time temporal systems is the \emph{dynamic Bayesian network (DBN)} model \citep{murphyMachineLearningProbabilistic2012}, which captures relationships between variables across time.

A DBN is specified by a pair of DAGs, and a corresponding pair of factorized distributions.
The \emph{prior network} ${\cal G}_1$ and its corresponding distribution $p({\bf V}_1) = \prod_{V_i \in {\bf V}_1} p(V_i \mid \pa_{{\cal G}_1}(V_i))$ represent the state of the system at the first time step.  The \emph{transition network} ${\cal G}_{t,t+1}$ is a \emph{conditional DAG (CDAG)} with random vertices ${\bf V}_{t+1}$ representing variables at time point $t+1$, and fixed vertices ${\bf V}_t$ representing context in the previous time point $t$.  We will describe such conditional graphs by a shorthand ``${\cal G}_{t+1,t}$ on ${\bf V}_{t+1}$ given ${\bf V}_t$.''
In this conditional DAG no arrowheads into vertices in ${\bf V}_t$ are allowed.  The corresponding conditional distribution $p({\bf V}_{t+1} \mid {\bf V}_t)$ represents the way variables at point $t+1$ depend on each other, and on variables at the prior time point $t$ (and on no other prior variables, such as those at time point $t-1$).   This dependence leads to a \emph{first order Markov} DBN.
This distribution factorizes with respect to the CDAG ${\cal G}_{t,t+1}$ as follows:
$p({\bf V}_{t+1} \mid {\bf V}_t) = \prod_{V_i \in {\bf V}_{t+1}} p(V_i \mid \pa_{{\cal G}_{t,t+1}}(V_i))$.

The joint distribution for the DBN system over a finite number of discrete time steps $T$ is given by 
the product of the prior network distribution, and the transition conditional probability distributions for a set of time steps, as follows: 

\vspace{-0.55cm}
{\small 
	\begin{align}
	\left( \prod_{V \in {\bf V}_1} p(V | \pa_{\G_1}(V))\! \right) \!\cdot \!
	\prod_{t=1}^{T-1}
	\!\!\left(\! \prod_{V \in {\bf V}_{t+1}} p(V | \pa_{\G_{t,t+1}}(V)) \right) \numberthis
	\label{eqn:dag-lld}
	\end{align}
}
\vspace{-0.4cm}

A simple DBN is represented in Figure ~\ref{fig:dynamic}, where the prior network (~\ref{fig:dynamic}(a)) contains two variables $A$ and $L$, and the transition network (~\ref{fig:dynamic}(b)) shows connections among the state variables in the prior state at time $t$ and the subsequent state at time $t+1$.  We represent fixed vertices in a transition network via squares.
Figure ~\ref{fig:dynamic}(c) shows the DBN implied by these prior and transition networks unrolled over $4$ time steps. DBNs can be naturally extended to represent causal models by assuming that both prior and transition networks are causal DAGs. 
In other words, we assume values of every variable $V_i$ in both the prior and the transition network is determined, via a structural equation $f_i(.)$, in terms of its observed parents $\pa_{{\cal G}_1}(V_i)$ (or $\pa_{{\cal G}_{t,t+1}}(V_i)$) and an exogenous noise term $\epsilon_i$.  If we further assume that all exogenous noise variables are marginally independent, we arrive at a DBN version of the NPSEM-IE, where in addition to the g-formula (\ref{eqn:g}) holding for the prior network, the \emph{conditional g-formula} holds for the transition network:

\vspace{-0.55cm}
{\small
\begin{align}
p({\bf Y}_{t+1}({\bf a}) | {\bf V}_t) =
\!\!\!\!
\prod_{V_i \in {\bf Y}_{t+1}} p(V_i | \pa_{\cal G}(V_i))
\vert_{ {\bf A} = {\bf a} },
\label{eqn:c-g}
\end{align}
}
\vspace{-0.5cm}

for any ${\bf A} \subseteq {\bf V}_{t+1}$, and ${\bf Y}_{t+1} = {\bf V}_{t+1} \setminus {\bf A}$.

Thus, a causal DBN ``unrolled'' to a set of time points $1, \ldots, T$ yields a standard causal DAG model with vertices ${\bf V}_{1:T} \equiv {\bf V}_1 \cup {\bf V}_2 \cup \ldots \cup {\bf V}_T$ .  In particular, for an intervention that sets ${\bf A} \subseteq {\bf V}_{1:T}$ to constant values ${\bf a}$, the interventional distribution $p({\bf Y}_{1:T}({\bf a}))$, where
${\bf Y}_{1:T} = {\bf V}_{1:T} \setminus {\bf A}$, is identified by:

\vspace{-0.5cm}
{\small 
\begin{equation}
\prod_{V \in {\bf V}_1 \setminus {\bf A}} \!\!\!\! p(V | \pa_{\G_1}(V)) 
\prod_{t=1}^{T-1}
\prod_{V \in {\bf V}_{t+1}  \setminus {\bf A} } \!\!\!\! p(V | \pa_{\G_{t,t+1}}(V))
\Bigg\vert_{\mathrlap{{\bf A} = {\bf a}}}
\numberthis \label{eqn:g-dbn}
\end{equation}
}
\vspace{-0.5cm}

Causal DBNs have been considered in prior work. \cite{petersCausalInferenceTime2013} illustrated how structural equations can be used in the context of time series data, addressing issues of identifiability. \cite{malinskyCausalStructureLearning2018, malinskyLearningStructureNonstationary2019, mogensen2018causal} presented structure learning algorithms for causal dynamic networks and applied them to macroeconomic data.

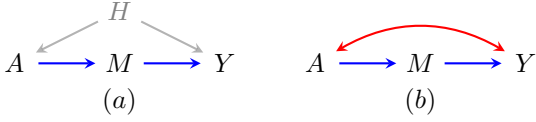
\begin{figure}[t]
	\begin{center}
		\begin{tikzpicture}[>=stealth, node distance=1.4cm]
		\tikzstyle{format} = [draw=none, very thick, circle, minimum size=6mm,
		inner sep=0pt]
		\tikzstyle{square} = [draw, very thick, rectangle, minimum size=5.0mm, inner sep=0pt]

		\begin{scope}[xshift=-4cm, yshift=0cm]
		\path[->, thick]
		node[format] (a) {$A$}
		node[format, right of=a] (m) {$M$}
		node[format, above of=m, yshift=-0.7cm, opacity=0.5] (h) {$H$}
		node[format, right of=m] (y) {$Y$}
		
		(a) edge[blue] (m)
		(m) edge[blue] (y)
		(h) edge[black,opacity=0.3] (a)
		(h) edge[black,opacity=0.3] (y)
		
		node[below of=m, yshift=0.9cm, xshift=0cm] (l) {$(a)$}
		;
		\end{scope}
		
		\begin{scope}[xshift=0cm, yshift=0cm]
		\path[->, thick]
		node[format] (a) {$A$}
		node[format, right of=a] (m) {$M$}
		node[format, right of=m] (y) {$Y$}
		
		(a) edge[blue] (m)
		(m) edge[blue] (y)
		(a) edge[<->, red, bend left] (y)
		
		node[below of=m, yshift=0.9cm, xshift=0cm] (l) {$(b)$}
		;
		\end{scope}

		\end{tikzpicture}
	\end{center}
	\caption {A hidden variable DAG and latent projection ADMG
	}
	\label{fig:front-door}
\end{figure}

\subsection{Causal Inference With Hidden Variables}
\label{sec:admg}
The g-formula (\ref{eqn:g}) provides an elegant link between observed data and counterfactual distributions in causal models where all relevant variables are observed.
Causal models that arise in practice, however,  
contain hidden variables.  
Representing such models using a DAG ${\cal G}({\bf V} \cup {\bf H})$ where 
${\bf V}$ and ${\bf H}$ correspond to observed and hidden variables, respectively,
is not very helpful, since 
applying (\ref{eqn:g}) to ${\cal G}({\bf V} \cup {\bf H})$ results in an expression that involves unobserved variables ${\bf H}$.

A popular alternative is to represent a class of hidden variable DAGs ${\cal G}_i({\bf V} \cup {\bf H}_i)$ by a single
\emph{acyclic directed mixed graph} ADMG ${\cal G}({\bf V})$ that contains directed ($\to$) and bidirected ($\leftrightarrow$) edges and no directed cycles
via the \emph{latent projection} operation \citep{verma90equiv} (see Section \ref{nested-markov} of the Appendix).
The latent projection ADMG ${\cal G}({\bf V})$ captures relationships between observed variables ${\bf V}$ implied by the factorization of $p({\bf V} \cup {\bf H})$ with respect to ${\cal G}({\bf V} \cup {\bf H})$ via the \emph{nested Markov factorization} of $p({\bf V})$ with respect to ${\cal G}({\bf V})$ \citep{richardson17nested}.

In particular, just as identification in DAGs may be viewed in terms of a modified DAG factorization (\ref{eqn:g}), identification in a hidden variable DAG ${\cal G}({\bf V} \cup {\bf H})$ may be viewed in terms of a modified nested factorization of ${\cal G}({\bf V})$. 
The {nested Markov factorization} of 
$p({\bf V})$ with respect to 
${\cal G}({\bf V})$ is defined in terms of \emph{Markov kernels} of the form $q_{\bf D}({\bf D} | \pa_{\cal G}({\bf D}) \setminus {\bf D})$, where set ${\bf D} \subseteq {\bf V}$ is \emph{intrinsic} in ${\cal G}({\bf V})$ (see below).
Kernels $q_{\bf D}({\bf D} | \pa_{\cal G}({\bf D}) \setminus {\bf D})$ are objects that resemble conditional densities $p(V_i \mid \pa_{\cal G}(V_i))$ that arise in the
Markov factorization for a DAG, in the sense that they are non-negative and normalize to $1$ for every value of $\pa_{\cal G}({\bf D}) \setminus {\bf D}$.
Kernels making up the nested Markov factorization are all functionals of $p({\bf V})$, however, they are not necessarily equal to conditional distributions $p({\bf D} \mid \pa_{\cal G}({\bf D}) \setminus {\bf D})$.

A set ${\bf D}$ is intrinsic if it is bidirected connected and reachable.  A set ${\bf D}$ is reachable if it is possible to find an order on variables $\langle S_1, S_2, \ldots, S_m \rangle \equiv {\bf V} \setminus {\bf D}$ such that in each subgraph ${\cal G}_i({\bf V})$, obtained from ${\cal G}({\bf V})$ by removing all vertices $\{ S_1, \ldots, S_{i-1} \}$ and adjacent edges, there is no variable $W$ that is a descendant of $S_i$ in ${\cal G}_i({\bf V})$, and simultaneously has a path to $S_i$ consisting exclusively of bidirected edges. 

The nested Markov factorization asserts that the observed  
margin $p({\bf V})$ can be expressed as a product 
$\prod_{{\bf D} \in {\cal D}({\cal G}({\bf V}))} q_{\bf D}({\bf D} \mid \pa_{\cal G}({\bf D}) \setminus {\bf D})$ of kernels where ${\cal D}({\cal G}({\bf V}))$ is the set of bidirected connected components, called \emph{districts}, in ${\cal G}({\bf V})$.  Additionally, the factorization implies certain other kernels associated with reachable sets may be expressed as similar products of intrinsic kernels.   Finally, the modified form of the factorization may be used to express \emph{any} interventional distribution identified from $p({\bf V})$, as follows.

Given a latent projection ADMG ${\cal G}({\bf V})$ representing a hidden variable causal model, and
any disjoint subsets ${\bf Y},{\bf A}$ of ${\bf V}$, let ${\bf Y}^*$ be the set of ancestors of ${\bf Y}$ in
${\cal G}({\bf V})$ via directed paths that do not pass through ${\bf A}$, and let ${\cal G}_{{\bf Y}^*}$
be the \emph{induced subgraph} of ${\cal G}({\bf V})$ containing only vertices in ${\bf Y}^*$ and edges among these vertices.

\citep{shpitser06id,richardson17nested} showed that any interventional distribution $p({\bf Y}({\bf a}))$ 
is identified from $p({\bf V})$ given ${\cal G}({\bf V})$ if and only if every bidirected connected component in ${\cal G}_{{\bf Y}^*}$ is intrinsic.
Moreover, if $p({\bf Y}({\bf a}))$ is identified, it is given by the following margin of the modified nested Markov factorization, made up of the appropriate kernels:

\vspace{-0.5cm}
{\small
	\setlength{\abovedisplayskip}{5pt} \setlength{\belowdisplayskip}{5pt}
	\begin{align}
	p({\bf Y}({\bf a})) = 
	\sum_{{\bf Y}^* \setminus ({\bf Y} \cup {\bf A})} \prod_{{\bf D} \in {\cal D}({\cal G}_{{\bf Y}^*})} q_{\bf D}({\bf D} | \pa_{\cal G}({\bf D}) \setminus {\bf D}) \vert_{{\bf A} = {\bf a}}.
	\label{eqn:id}
	\end{align}
}
\vspace{-0.4cm}

As a simple example, consider the hidden variable DAG in Fig.~\ref{fig:front-door}(a). Its latent projection ADMG in Fig.~\ref{fig:front-door}(b), called the \emph{front-door graph}, has intrinsic sets $\{ A \}, \{ M \}, \{ A, Y \}, \{ Y \}$, with the corresponding kernels:
$q_A(A) \equiv p(A)$, $q_M(M | A) \equiv p(M | A)$, $q_{A,Y}(A,Y | M) \equiv p(Y | M, A) p(A)$, and $q_{Y}(Y \mid M) \equiv \sum_{A} p(Y | M, A) p(A)$.

By the nested Markov factorization, 
the observed margin $p(A,M,Y)$ factorizes as $q_{A,Y}(A,Y | M) q_{M}(M | A)$.  In addition, certain other distributions also factorize.  For example, the margin $p(A,M)$ is equal to $q_A(A) q_M(M | A)$.  
Further, $p(Y(a))$ is identified from $p(A,M,Y)$ and equal to $\sum_{M} q_Y(Y | M) q_M(M | a) = \sum_M \left( \sum_{A'} p(Y | M,A') p(A') \right) p(M | a)$, which is the \emph{front-door formula} \citep{pearl95causal}. See Section \ref{nested-markov} of the Appendix for a detailed exposition of the nested Markov factorization and identification theory in ADMGs. 

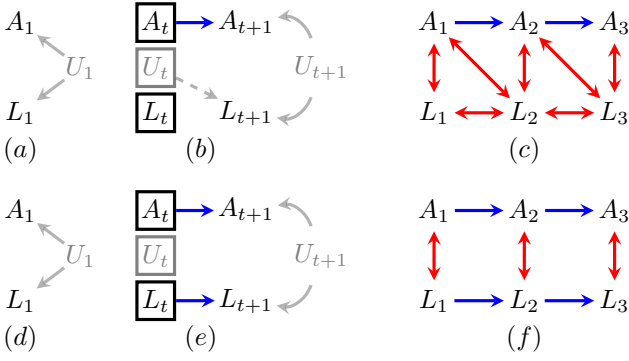
\begin{figure}  %
	\begin{center}
		\begin{tikzpicture}[>=stealth, node distance=1.2cm]
		\tikzstyle{format} = [draw=none, very thick, circle, minimum size=5.0mm,
		inner sep=0pt]
		\tikzstyle{square} = [draw, very thick, rectangle, minimum size=5.0mm, inner sep=0pt]
		
		\begin{scope}[xshift=-3cm,yshift=0cm]
		\path[->, very thick]
		node[format] (a) {$A_1$}
		node[format, below of=a] (l) {$L_1$}
		node[format, below of=a, xshift=0.8cm,yshift=0.6cm, opacity=0.5] (u) {$U_1$}
		
		(u) edge[black,opacity=0.3] (a)
		(u) edge[black,opacity=0.3] (l)

		node[below of=l, yshift=0.7cm, xshift=0.0cm] (l) {$(a)$}
		;
		\end{scope}
		
		\begin{scope}[xshift=-1.2cm,yshift=0cm]
		\path[->, very thick]
		node[square] (a) {$A_t$}
		node[square, below of=a] (l) {$L_t$}
		node[square, below of=a,xshift=-0cm,yshift=0.6cm, opacity=0.5] (u) {$U_t$}
		
		node[format, right of=a] (a2) {$A_{t+1}$}
		node[format, below of=a2] (l2) {$L_{t+1}$}
		node[format, below of=a2,xshift=1cm,yshift=0.6cm, opacity=0.5] (u2) {$U_{t+1}$}
		
		(a) edge[blue] (a2)
		(u) edge[black, opacity=0.3, dashed, bend right =0] (l2)
		(u2) edge[black, opacity=0.3, bend right=40] (a2)
		(u2) edge[black, opacity=0.3, bend left=40] (l2)
		
		node[below of=l, yshift=0.7cm, xshift=.6cm] (l) {$(b)$}
		;
		\end{scope}
		
		\begin{scope}[xshift=2.5cm,yshift=0cm]
		\path[->, very thick]
		node[format] (a0) {$A_1$}
		node[format, below of=a0] (l0) {$L_1$}
		
		node[format, right of=a0] (a1) {$A_{2}$}
		node[format, below of=a1] (l1) {$L_{2}$}
		
		node[format, right of=a1] (a2) {$A_{3}$}
		node[format, below of=a2] (l2) {$L_{3}$}
		
		(a0) edge[<->,red] (l1)
		(a1) edge[<->,red] (l2)
		
		(a0) edge[blue] (a1)
		(a1) edge[blue] (a2)
		(a0) edge[<->,red] (l0)
		(a1) edge[<->,red] (l1)
		(a2) edge[<->,red] (l2)
		(l0) edge[<->,red] (l1)
		(l1) edge[<->,red] (l2)
		
		node[below of=l1, yshift=0.7cm, xshift=0cm] (l) {$(c)$}
		;
		\end{scope}

		\begin{scope}[xshift=-3cm,yshift=-2.5cm]
		\path[->, very thick]
		node[format] (a) {$A_1$}
		node[format, below of=a] (l) {$L_1$}
		node[format, below of=a, xshift=0.8cm,yshift=0.6cm, opacity=0.5] (u) {$U_1$}
		
		(u) edge[black,opacity=0.3] (a)
		(u) edge[black,opacity=0.3] (l)

		node[below of=l, yshift=0.7cm, xshift=0.0cm] (l) {$(d)$}
		;
		\end{scope}
		
		\begin{scope}[xshift=-1.2cm,yshift=-2.5cm]
		\path[->, very thick]
		node[square] (a) {$A_t$}
		node[square, below of=a] (l) {$L_t$}
		node[square, below of=a,xshift=-0cm,yshift=0.6cm, opacity=0.5] (u) {$U_t$}
		
		node[format, right of=a] (a2) {$A_{t+1}$}
		node[format, below of=a2] (l2) {$L_{t+1}$}
		node[format, below of=a2,xshift=1cm,yshift=0.6cm, opacity=0.5] (u2) {$U_{t+1}$}
		
		(a) edge[blue] (a2)
		(l) edge[blue, bend right =0] (l2)
		(u2) edge[black, opacity=0.3, bend right=40] (a2)
		(u2) edge[black, opacity=0.3, bend left=40] (l2)
		
		node[below of=l, yshift=0.7cm, xshift=.6cm] (l) {$(e)$}
		;
		\end{scope}
		
		\begin{scope}[xshift=2.5cm,yshift=-2.5cm]
		\path[->, very thick]
		node[format] (a0) {$A_1$}
		node[format, below of=a0] (l0) {$L_1$}
		
		node[format, right of=a0] (a1) {$A_{2}$}
		node[format, below of=a1] (l1) {$L_{2}$}
		
		node[format, right of=a1] (a2) {$A_{3}$}
		node[format, below of=a2] (l2) {$L_{3}$}
		
		(a0) edge[blue] (a1)
		(a1) edge[blue] (a2)
		(a0) edge[<->,red] (l0)
		(a1) edge[<->,red] (l1)
		(a2) edge[<->,red] (l2)

		(l0) edge[->,blue] (l1)
		(l1) edge[->,blue] (l2)
		
		node[below of=l1, yshift=0.7cm, xshift=0cm] (l) {$(f)$}
		;
		\end{scope}		
		
		\end{tikzpicture}
	\end{center}
	\caption{
		(a),(d) Prior network hidden variable DAGs $\G_0$, representing the state at time $t=0$.
		(b),(e) Conditional hidden variable DAGs $\G_{t,t+1}$ representing the transitions in the network, with (e) leading to a first-order Markov model, and (b) leading to
			higher order dependences to unobserved hidden variables $U_t$ linking multiple time points.
		(c),(f) Latent projection ADMGs of the unrolled hidden variable DBNs to three time steps.
	}
	\label{fig:causal-dynamic-hidden}
\end{figure}

\section{Identification In Hidden Variable DBNs}
\label{sec:temporal-hidden-vars}

Identification theory in hidden variable causal models can be extended to 
hidden variable causal DBN models as well. To the best of our knowledge, this extension is novel.

We start with an assumption that will allow us to view the marginal version of the DBN, defined only on observed variables,
as a first-order Markov DBN.
\begin{assump}\label{assump:fixed-observed}
	The transition network ${\cal G}_{t+1,t}$ only depends on fixed variables in the previous time step $t$ that are observed.
\end{assump}
If ${\cal G}_{t+1,t}$
depends on fixed variables that are hidden, the resulting DBN may result in observed variables in step $t+1$ that depend on observed variables earlier than $t$ even if observed variables in $t$ are conditioned on.

For example, consider the DBN specified by prior and transition networks in Fig.~\ref{fig:causal-dynamic-hidden} (a) and (b).  Because the variable $L_{t+1}$ depends on $U_t$, which is unobserved, and $U_t$ influences $L_t$, ``unrolling'' this network, and taking the latent projection yields an ADMG shown in Fig.~\ref{fig:causal-dynamic-hidden} (c), where 
$L_3$ ends up being dependent on $L_1$, even after conditioning on $L_2,A_2$ 
(due to the ``explaining away'' phenomenon arising when a shared effect $L_2$ of two variables $U_2$ and $U_1$ is conditioned on).  On the other hand, the DBN specified by prior and transition networks in Fig.~\ref{fig:causal-dynamic-hidden} (d) and (e) does not suffer from this issue, as the transition network only depends on observed variables $L_t,A_t$, yielding a latent projection of the ``unrolled'' model shown in Fig.~\ref{fig:causal-dynamic-hidden} (f), which factorizes into time step specific conditional distributions: $p(A_1, L_1) p(A_2, L_2 | A_1, L_1) p(A_3, L_3 | A_2, L_2)$.

In general, given a hidden variable prior network ${\cal G}_1$ on ${\bf V}_1, {\bf H}_1$, and transition network ${\cal G}_{t+1,t}$ on ${\bf V}_{t+1}, {\bf H}_{t+1}$ given ${\bf V}_t$, the hidden variable DBN may be represented by latent projections of the prior and transition networks: an ADMG ${\cal G}_1$ on ${\bf V}_1$, and a \emph{conditional ADMG (CADMG)} ${\cal G}_{t+1,t}$ on ${\bf V}_{t+1}$ given ${\bf V}_t$, and the corresponding marginal distributions $p({\bf V}_1)$ and $p({\bf V}_{t+1,t} | {\bf V}_t)$.
The ``unrolled'' version of the factorization of this model is:
$p({\bf V}_1) \prod_{t=1}^T p({\bf V}_{t+1,t} | {\bf V}_t)$,
where each term nested Markov factorizes with respect to either ${\cal G}_1$ or ${\cal G}_{t+1,t}$ by results in \citep{richardson17nested}.
Intrinsic and reachable sets are defined in CADMGs by ignoring fixed vertices, and the nested factorization generalizes in the natural way from ADMGs from CADMGs, see Section \ref{nested-markov} of the Appendix for details.

If the underlying DAGs correspond to causal models, the hidden variable DBN yields identification theory where the modified nested factorization (\ref{eqn:id}) is applied at every time point, just as (\ref{eqn:g}) was applied at every point in a fully observed causal DBN to yield (\ref{eqn:g-dbn}).
That is, given a fixed set of time points $1, \ldots, T$, vertices ${\bf V}_{1:T} \equiv {\bf V}_1 \cup {\bf V}_2 \cup \ldots \cup {\bf V}_T$, and
disjoint subsets ${\bf A},{\bf Y} \subseteq {\bf V}_{1:T}$, we have the following result: 

\begin{lem}
\label{eqn:hidden-var-causal-dbn}
Under Assumption \ref{assump:fixed-observed} , $p({\bf Y}({\bf a}))$ is identified from a hidden variable causal DBN model represented by latent projections ${\cal G}_1$ on ${\bf V}_1$ and ${\cal G}_{t+1,t}$ on ${\bf V}_{t+1}$ given ${\bf V}_t$ if and only if every bidirected connected component in ${{\cal G}_1}_{{\bf Y}^*_1}$ (the induced subgraph of ${\cal G}_1$) is intrinsic in ${\cal G}_1$, and every bidirected component in ${{\cal G}_{t+1,t}}_{{\bf Y}^*_{i}}$ (the induced subgraph of ${\cal G}_{t+1,t}$) is intrinsic in ${\cal G}_{t+1,t}$, where ${\bf Y}^*_1$ is the set of ancestors of ${\bf Y} \cap {\bf V}_1$ not through ${\bf A} \cap {\bf V}_1$ in ${\cal G}_1$, and for every $i \in 2,\ldots, T$, ${\bf Y}^*_{i}$ is the set of ancestors of ${\bf Y} \cap {\bf V}_{i}$ not through ${\bf A} \cap {\bf V}_{i}$ in ${\cal G}_{t+1,t}$.  Moreover, if 
$p({\bf Y}({\bf a}))$ is identified, we have
{\small
\begin{align*}
\Bigg( \sum_{{\bf Y}^*_1 \setminus (({\bf Y} \cup {\bf A}) \cap {\bf V}_1)} 
	 \prod_{{\bf D} \in {\cal D}({{\cal G}_1}_{{\bf Y}^*_1})} q^1_{\bf D}({\bf D} | \pa_{\cal G}({\bf D}) \setminus {\bf D}) \vert_{{\bf A} = {\bf a}}
\Bigg) \times\\
\prod_{i=2}^T
\Bigg( \sum_{{\bf Y}^*_i \setminus (({\bf Y} \cup {\bf A}) \cap {\bf V}_i)} 
	 \prod_{{\bf D} \in {\cal D}({{\cal G}_{t+1,t}}_{{\bf Y}^*_i})} q^{t+1,t}_{\bf D}({\bf D} | \pa_{\cal G}({\bf D}) \setminus {\bf D}) \vert_{{\bf A} = {\bf a}},
\Bigg)
\end{align*}
}
\noindent
where $q^1_{\bf D}$ and $q^{t+1,t}_{\bf D}$ are kernels corresponding to intrinsic sets 
that are districts in ${\cal D}({{\cal G}_1}_{{\bf Y}^*_1})$ and ${\cal D}({{\cal G}_{t+1,t}}_{{\bf Y}^*_1})$ in the nested Markov factorizations of ${\cal G}_1$ and ${\cal G}_{t+1,t}$, respectively.
\end{lem}
We present the proof of this result and a worked example in the Section \ref{nested-markov} of the Appendix.

\section{Fully Observed PDSEMs}\label{pdsem-dags}
A crucial modeling assumption employed by 
causal DBNs is that both structure and parameterization remain invariant over time. Such a model is ill-suited to capture the sort of path dependence described in the introduction. We now describe our new approach to relaxing this assumption via path dependent structural equation models (PDSEMs), which are a generalization of causal DBNs that can capture path dependence. 
\subsection{A Simple PDSEM} \label{pdsem-dags-toy}
To illustrate PDSEMs, we will use a simple example inspired by our surgery setting.
We assume a surgery will consist of three states: $s^1$ (``incision''), 
$s^2$ (``modification of bone/tissue''), and $s^3$ (``closing the incision'').
Further, each state has the following variables: $A$ (patient status prior to any procedures in the current stage), $B$ (experience of surgeon performing the procedure in the current stage)
and $C$ (the observed patient outcome for the stage after procedure is performed), all observed. The surgery always starts at $s^1$, and concludes upon reaching $s^3$.  Procedures performed in $s^2$ may either succeed, leading to $s^3$, or fail with some probability, leading the surgeon to revisit $s^1$.
Note that DBNs are unable to capture even this simple example, due to the fact that DBN cannot represent distinct transitions to qualitatively different states. The state transition diagram for this scenario is shown in Fig.~\ref{fig:dags-sim} (b).  By contrast, the only type of state transition diagram allowed by DBNs contains a single self-looping state.

Relationships between variables in $s^1$ 
are shown by a causal diagram in Fig.~\ref{fig:dags-sim} (a), corresponding to a set of structural equations as described in Sec \ref{sec:dag-model}. Fig.~\ref{fig:dags-sim} (a) 
serves the role played by the prior network in a causal DBN. In addition to variables $A_1, B_1$ and $C_1$, this graph contains variable $S_1$, representing the state to transition to, at time step $1$.  In general, the probability associated with this variable may depend on other variables in the current state, however in our simple model, the state $s^1$ transitions to $s^2$ with probability $1$.

Transitions are specified by causal CDAGs for each possible state transition, shown in Fig.~\ref{fig:dags-sim}(c),(d) and (e) (without the dashed edge). 
These graphs include edges between variables in time steps ${t}$ and $t+1$, as well as state-specific relationships among variables at time $t+1$. Note that unlike a DBN, which only had a single transition network, multiple transition networks are needed to represent multiple transitions between different states.
We assume state spaces of variables associated with each state are the same across 
state transition and prior graphs, though variables themselves and their causal relationships may differ across graphs.
For example, the state spaces of $A_1,B_1,C_1$ in Fig.~\ref{fig:dags-sim}(a) and $A_{21},B_{21},C_{21}$ in Fig.~\ref{fig:dags-sim}(c) are the same, while the variables themselves (and the causal graphs relating them) are not.  This implies values may be indexed by state, e.g. $a_1$ can refer without loss of generality to a value of $A_1$ or $A_{21}$.
We thus can thus index conditional distributions that depend on a prior state only by the prior state itself, e.g. $p(A_{12} | A_1)$ is a shorthand for ``a density over $A_{12}$ in transition $(1,2)$ given any value $a_1$ of any variable of the form $A_{i1}$.''

Causal graphs in \ref{fig:dags-sim}(a),(c),(d),(e), along with the state-transition diagram \ref{fig:dags-sim}(b), completely describe the fully observed PDSEM. Complex state dynamics are captured by distinct state causal DAGs and path-dependence is simply a consequence of state-transitions that depend on variables in the current state, and not just the state itself. 

\begin{figure}[t]
	\begin{center}
		\begin{tikzpicture}[>=stealth, node distance=1.4cm]
		\tikzstyle{format} = [draw=none, very thick, circle, minimum size=6mm,
		inner sep=0pt]
		\tikzstyle{square} = [draw, very thick, rectangle, minimum size=5.0mm, inner sep=0pt]

		\begin{scope}[xshift=0cm, yshift=-2.5cm]
		\path[->, thick]
		node[format] (a) {$A_1$}
		node[format, below left of=a](b) {$B_1$}
		node[format, below right of=a](c) {$C_1$}
		node[format, right of=a, xshift=-0.4cm] (s) {$S_1$}
		
		(a) edge[blue] (c)
		(b) edge[blue] (c)		
		
		node[below of=a, yshift=-0.3cm, xshift=0.0cm] (la) {$(a) \: s^1$}
		;
		\end{scope}
		
		\begin{scope}[xshift=3cm,yshift=-2.5cm]
		\path[->, thick]
		node[format] (a) {$s^{1}$}
		node[format, below left of=a,,xshift=0cm,yshift=-0.0cm] (b) {$s^{2}$}
		node[format, below right of=a,xshift=0cm,yshift=-0.0cm] (c) {$s^{3}$}
		
		(b) edge[black, bend left=30] (a)
		(a) edge[black, bend left=30] (b)
		(b) edge[black] (c)		
		(c) edge[loop, in=60,out=130,looseness=5] (c)
		
		node[below of=a, yshift=-0.3cm, xshift=0.0cm] (la) {$(b)$}
		;
		\end{scope}
		
		\begin{scope}[yshift=-4.7cm, xshift=-0.8cm]
		\path[->, thick]
		node[square] (a1) {$A_{1}$}
		node[square, below of=a1] (b1) {$B_{1}$}
		node[square, below of=b1] (c1) {$C_{1}$}
		node[format, right of=a1, xshift=-0.3cm] (a2) {$A_{12}$}
		node[format, below of=a2, xshift=-0.0cm] (b2) {$B_{12}$}
		node[format, below of=b2, xshift=-0.0cm] (c2) {$C_{12}$}
		node[format, right of=c2, xshift=-0.4cm] (s) {$S_{12}$}

		(b2) edge[blue] (c2)
		(a2) edge[blue, bend left] (c2)

		(c2) edge[blue] (s)
		
		(a1) edge[blue] (a2)
		(c1) edge[blue] (a2)
		(c1) edge[blue] (c2)
		(a2) edge[<->, red, dashed] (b2)
		
		node[below of=a1, yshift=-2.0cm, xshift=0.6cm] (la) {$(c) \: \mscriptsize{s^1_{t} \rightarrow s^2_{t+1}}$}
		;
		\end{scope}
		
		\begin{scope}[yshift=-4.7cm, xshift=2.0cm]
		\path[->, thick]
		node[square] (a1) {$A_{2}$}
		node[square, below of=a1] (b1) {$B_{2}$}
		node[square, below of=b1] (c1) {$C_{2}$}
		node[format, right of=a1, xshift=-0.3cm] (a2) {$A_{23}$}
		node[format, below of=a2] (b2) {$B_{23}$}
		node[format, below of=b2] (c2) {$C_{23}$}
		node[format, right of=c2, xshift=-0.6cm] (s) {$S_{23}$}
		
		
		(a1) edge[blue] (a2)
		(b1) edge[blue] (b2)
		(c1) edge[blue] (c2)
		(c1) edge[blue] (a2)

		(a2) edge[blue] (b2)
		(a2) edge[blue, bend left] (c2)
		
		node[below of=a1, yshift=-2.0cm, xshift=0.6cm] (la) {$(d) \:\mscriptsize{ s^2_{t} \rightarrow s^3_{t+1}}$}
		;
		\end{scope}

		\begin{scope}[yshift=-4.7cm, xshift=4.6cm]
		\path[->, thick]
		node[square] (a1) {$A_{2}$}
		node[square, below of=a1] (b1) {$B_{2}$}
		node[square, below of=b1] (c1) {$C_{2}$}
		node[format, right of=a1, xshift=-0.3cm] (a2) {$A_{21}$}
		node[format, below of=a2] (b2) {$B_{21}$}
		node[format, below of=b2] (c2) {$C_{21}$}

		node[format, right of=c2, xshift=-0.6cm] (s) {$S_{21}$}
		
		
		(a1) edge[blue] (a2)
		(b1) edge[blue] (b2)
		(c1) edge[blue] (c2)
		(c1) edge[blue] (a2)
		(a1) edge[blue] (b2)
		(c1) edge[blue] (b2)

		(b2) edge[blue] (c2)
		(a2) edge[blue, bend left] (c2)

		node[below of=a1, yshift=-2.0cm, xshift=0.6cm] (la) {$(e) \: \mscriptsize{s^2_{t} \rightarrow s^1_{t+1}}$}
		;
		\end{scope}

		\begin{scope}[yshift=-9.3cm, xshift=-0.6cm]
		\path[->, thick]
		
		node[format, shape=rectangle, draw, color=black, opacity=0.3, dashed, minimum width=30mm, minimum height=6mm, xshift=1cm] (r1) {$ $}
		node[format] (s11) {$s^1$}
		node[format, right of=s11, xshift=-0.4cm] (s12) {$s^2$}
		node[format, right of=s12, xshift=-0.4cm] (s13) {$s^3$}

		node[format, below of=s11, yshift=0.7cm] (s21) {$s^1$}
		node[format, right of=s21, xshift=-0.4cm] (s22) {$s^2$}
		node[format, right of=s22, xshift=-0.4cm] (s23) {$s^3$}
		
		node[format, below of=s21, yshift=0.7cm] (s31) {$s^1$}
		node[format, right of=s31, xshift=-0.4cm] (s32) {$s^2$}
		node[format, right of=s32, xshift=-0.4cm] (s33) {$s^3$}
		
		node[format, below of=s31, yshift=0.7cm, xshift=-.1cm] (s41) {$.$}	
		node[format, right of=s41, xshift=-1.3cm] (s42) {$.$}
		node[format, right of=s41, xshift=-1.2cm] (s43) {$.$}
		
		node[format, below of=s41, yshift=1cm] (s51) {$.$}	
		node[format, right of=s51, xshift=-1.3cm] (s52) {$.$}
		node[format, right of=s51, xshift=-1.2cm] (s53) {$.$}	
		
		node[format, below of=s51, yshift=1cm] (s61) {$.$}	
		node[format, right of=s61, xshift=-1.3cm] (s62) {$.$}
		node[format, right of=s61, xshift=-1.2cm] (s63) {$.$}			
		
		(s11) edge[black] (s12)
		(s12) edge[black] (s13)
		(s12) edge[black] (s21)
		(s21) edge[black] (s22)
		(s22) edge[black] (s23)
		(s22) edge[black] (s31)
		(s31) edge[black] (s32)
		(s32) edge[black] (s33)
		(s32) edge[black] (s41)
		node[below of=s11, yshift=-2.0cm, xshift=0.6cm] (la) {$(f)$}
		;
		\end{scope}
		
		\begin{scope}[yshift=-9.3cm, xshift=2.7cm, node distance=1.2cm]
		\path[->, thick]
		
		node[format, shape=rectangle, draw, color=black, opacity=0.3, dashed, minimum width=42mm, minimum height=34mm, xshift=1.8cm, yshift=-1.4cm] (r2) {$ $}

		node[format, xshift=0.5cm] (a1) {$A_1$}
		node[format, right of=a1, xshift=-0.0cm] (a2) {$A_{12}$}
		node[format, right of=a2, xshift=-0.0cm] (a3) {$A_{23}$}

		node[format, below of=a1, yshift=0.2cm] (b1) {$B_1$}
		node[format, right of=b1, xshift=-0.0cm] (b2) {$B_{12}$}
		node[format, right of=b2, xshift=-0.0cm] (b3) {$B_{23}$}
		
		node[format, below of=b1, yshift=0.2cm] (c1) {$C_1$}
		node[format, right of=c1, xshift=-0.0cm] (c2) {$C_{12}$}
		node[format, right of=c2, xshift=-0.0cm] (c3) {$C_{23}$}
		
		node[format, below right of=c1, xshift=-0.2cm,yshift=0.3cm] (S1) {$S_1$}
		node[format, below right of=c2, xshift=-0.2cm,yshift=0.3cm] (S2) {$S_2$}
		node[format, below right of=c3, xshift=-0.2cm,yshift=0.3cm] (S3) {$S_3$}
		
		(r1) edge[black, opacity=0.3,dashed, bend left=40] (r2)
%
		(a1) edge[blue] (b1)
		(b1) edge[blue] (c1)
		(a1) edge[blue, bend right=40] (c1)
		
		(a2) edge[<->,red, dashed, bend right=65] (b2)
		(b2) edge[blue] (c2)
		(a2) edge[blue, bend left=40] (c2)
		
		(a3) edge[blue] (b3)
		(a3) edge[blue, bend left=40](c3)

		(a1) edge[blue] (a2)
		(a2) edge[blue] (a3)
		(b2) edge[blue] (b3)
		(c1) edge[blue] (c2)
		(c2) edge[blue] (c3)
		
		(c1) edge[blue] (a2)
		(c2) edge[blue, bend left=5] (a3)
		
		(c2) edge[blue] (S2) 

		node[below of=a1, yshift=-2.3cm, xshift=1.6cm] (la) {$(g)$}
		;
		\end{scope}
	
		\end{tikzpicture}
	\end{center}
	\caption {A simple PDSEM.
	(a) Causal structure of the initial state $S^1$.
	(b) The state transition diagram.
	(c),(d),(e) Causal diagrams representing possible transitions and subsequent states.
	(f) Causal relationships in a system evolving according to the state transitions: $s^1 \to s^2 \to s^3$.
	(g) A snapshot of a possible PDSEM trajectory represented as an unrolled DAG}
	\label{fig:dags-sim}
\end{figure}
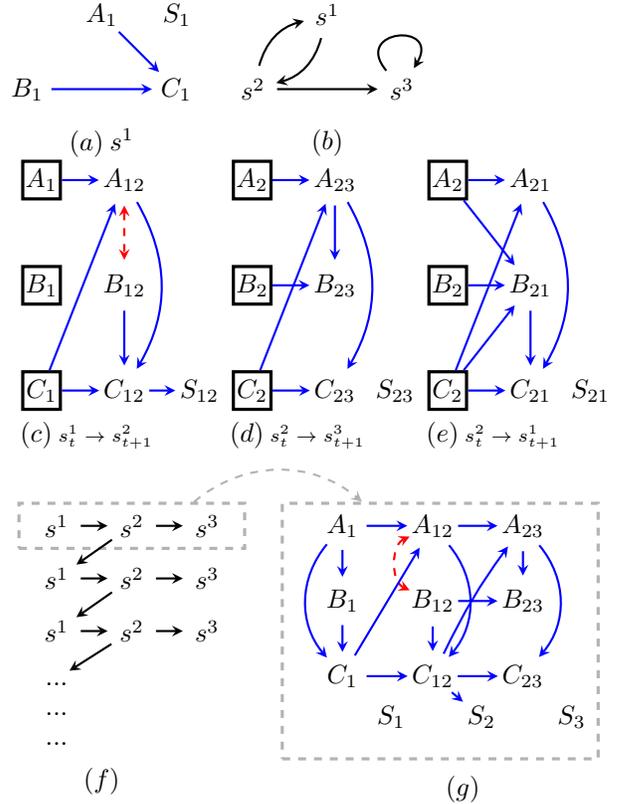

The model we describe represents a randomized controlled trial where the surgeon operating during state $s^2$ is randomly assigned, hence $B_{12}$ in the transition graph 
in Fig.~\ref{fig:dags-sim} (c) has no parents.  Otherwise, we encode standard causal relationships we expect: $C$ in the previous state influences $A,C$ in the next, and $A$ in the previous state influences $A$ in the next.  Surgeon assignment $B_{12}$ in $s^2$ influences assignments in subsequent stages, whether they are $s^1$ or $s^3$.  The state transition at $s^2$ depends on the outcome $C$ at that state.  In $s^3$, $B$ does not influence $C$, since closing the incision is a task adequately performed independent of surgeon experience.

The observed data factorization of a fully-observed PDSEM includes, in addition to factorization of the initial state causal DAG and factorizations with respect to CDAGs representing state transitions, state transition probabilities which are functions of variables in those factorizations. This factorization is not finite, but yields a well defined joint distribution $p_{\infty}$ over possible 
trajectories shown in \ref{fig:dags-sim}(f).
In our case, the distribution $p_{\infty}$ factorizes as follows:

\vspace{-0.5cm}
{\small
	\setlength{\abovedisplayskip}{5pt} \setlength{\belowdisplayskip}{5pt}
\begin{align*}
p_1 &\prod_{t=1}^{\infty}
\left(p_{12}\right)^{\mathbb{I}(s^1_{t}, s^2_{t+1})}
\left(p_{23}\right)^{\mathbb{I}(s^2_{t}, s^3_{t+1})}
\left(p_{21}\right)^{\mathbb{I}(s^2_{t}, s^1_{t+1})}
1^{\mathbb{I}(s^3_{t})}\\
p_1
\!
&=
\!
p(A_1) p(B_1 | A_1) p(C_1 | A_1,\! B_1) \tilde{p}(S_1)
\\
p_{12}
\!
&=
\!
p(A_{12} | A_{1}, \!C_{1}) p(B_{12})
p(C_{12} | B_{12},\! A_{12}, \!C_1) p(S_{12} | C_{12})
\\
p_{23}
\!
&=
\!
p(A_{23} | A_{2}, \!C_{2}) p(B_{23} | B_{2}, \!A_{23}) p(C_{23} | A_{23}, \!C_2) \tilde{p}(S_{23})
\\
p_{21}
\!
&=
\!
p(A_{21} | A_{2}, \!C_{2}) p(B_{21} | B_{2},\! A_{2},\! C_{2}) p(C_{21}  | C_{2}, \!B_{21},\! A_{21})
\tilde{p}(S_{21}),
\end{align*}
}
\vspace{-0.5cm}

where $s^i_{t}$ is the event ``the state at time $t$ is $s^i$, and all $\tilde{p}$ are deterministic by definition of our model.

PDSEMs, being causal models, allow us to reason about outcomes of interventions, for example questions such as:
``what would happen if all procedures at every stage of a surgery are performed by the resident surgeon ($B=b$), possibly contrary to fact.''
The counterfactual joint distribution $p_{\infty}(b)$ corresponding to this intervention is obtained by standard structural equation replacement semantics of interventions applied to the models representing initial and transition graphs \citep{pearl09causality}.
This distribution can be written as a product of state-specific marginal and conditional counterfactual distributions as follows:

\vspace{-0.5cm}
{\small
	\setlength{\abovedisplayskip}{3pt} \setlength{\belowdisplayskip}{3pt}
\begin{align*}
 p_1(b) \!\prod_{t=1}^{\infty}\!
\left(p_{12}(b)\right)^{\mathbb{I}(s^1_{t}, s^2_{t+1})}\!
\left(p_{23}(b)\right)^{\mathbb{I}(s^2_{t}), s^3_{t+1})}\!
\left(p_{21}(b)\right)^{\mathbb{I}(s^2_{t}, s^1_{t+1})}\!
1^{\mathbb{I}(s_{t}^3)}
\end{align*}
}
\vspace{-0.3cm}

The distribution $p_{\infty}(b)$ is identified by using the g-formula for every component of the factorization of $p_{\infty}$, in a generalization of
(\ref{eqn:g-dbn}), yielding:

\vspace{-0.6cm}
{\small
	\setlength{\abovedisplayskip}{5pt} \setlength{\belowdisplayskip}{5pt}
	\begin{align*}
&p_0^* \prod_{t=1}^{\infty}
\left(p_{12}^* \right)^{\mathbb{I}(s^1_{t}, s^2_{t+1})}
\left(p_{23}^* \right)^{\mathbb{I}(s^2_{t}), s^3_{t+1})}
\left(p_{21}^*\right)^{\mathbb{I}(s^2_{t}, s^1_{t+1})}
1^{\mathbb{I}(s_{t}^3)}
\\
& p_1^*=
p(A_1) p(C_1 | A_1, b) \tilde{p}(S_1)
\\
& p_{12}^*=
p(A_{12} | A_{1}, C_{1})
p(C_{12} | b, A_{12}, C_1) p(S_{12} | C_{12})
\\
& p_{23}^*=
p(A_{23} | A_{2}, C_{2}) p(C_{23} | A_{23}, C_2) \tilde{p}(S_{23})
\\
& p_{21}=
p(A_{21} | A_{2}, C_{2}) p(C_{21}  | C_{2}, b, A_{21}) \tilde{p}(S_{21}).
\end{align*}
}
While the distribution $p(S_{12} | C_{12})$ 
governing how likely it is that $s^1$ or $s^3$ are visited from $s^2$ remains the same, the probability that $s^1$ is visited from $s^2$ is likely higher in $p_{\infty}(b)$ compared to $p_{\infty}$.  This is because $B_{12}$ (counterfactually set to $b$) causes $C_{12}$, and $C_{12}$ causes $S_{12}$.
Thus, PDSEMs encode counterfactually changing state transition probabilities from their observed values.  
For example, surgeries where all stages are counterfactually performed by the less experienced resident surgeon might see more double-backs to $s^1$ to correct mistakes, compared to actually observed surgeries.

Having described this simple example, we turn to giving a general definition of PDSEMs.
\subsection{Arbitrary PDSEMs} 
An arbitrary PDSEM is defined using a set of states ${\bf s}$, with an initial state $s^1$ and an absorbing state $s^{*}$, a set ${\cal T}$ of state index pairs of the form $(i, j)$, where $s^i \neq s^*$ representing allowed state transitions, a DAG ${\cal G}_1$ on ${\bf V}_1$ for the initial state $s^1$, and for each $(i, j) \in {\cal T}$, a CDAG ${\cal G}_{ij}$ on ${\bf V}_{ij}$ given ${\bf V}_{i}$.
Variables $S_1 \in {\bf V}_1, \{ S_{ij} \in {\bf V}_{ij} : (i,j) \in {\cal T} \}$ determine probabilities of transitioning from state to state.
Just as in a causal DBN, the DAG ${\cal G}_1$, and CDAGs ${\cal G}_{ij}$ represent structural equation models for the initial state, and the appropriate state transitions, respectively.
That is, in the initial state, each variable $V \in {\bf V}_1$ is determined via $f_V(\pa_{\cal G}(V), \epsilon_V)$.  Similarly, for each variable $V \in {\bf V}_{ij}$ in any state transition represented by ${\cal G}_{ij}$.
We assume $S_1$, $\{ S_{ij} : (i,j) \in {\cal T} \}$ have no outgoing edges (this is without loss of generality, as structural equations are already state-specific in a PDSEM).

In order to formulate a first order Markov PDSEM, we need the following assumption that ensures that we need not condition on any context in the past except variables in the prior state.
\begin{assump}\label{assump:state-vars}
For every state $s^j$, any CDAG ${\cal G}_{ij}$ or DAG ${\cal G}_{j}$ will have random variables 
that share state spaces.
\end{assump}
We thus denote the values of any ${\bf V}_{ij}$ for any transition $(i,j)$ into state $j$ by ${\bf v}_j$ (note the lack of dependence on $i$).
As in our example above, we index conditional densities that depend on variables in a prior state by that state only, e.g.
$p(A_{12} | A_1)$.

Define ${\bf V} \equiv {\bf V}_1 \cup \left( \bigcup_{(i,j) \in {\cal T}} {\bf V}_{ij} \right)$. 
A PDSEM yields an observed data distribution $p_{\infty}({\bf V})$ with the following factorization: 
{\small
	\setlength{\abovedisplayskip}{3pt} \setlength{\belowdisplayskip}{3pt}
\begin{align*}
p_1({\bf V}_1) &\prod_{t=1}^{\infty} 
 \left( \prod_{(i,j) \in {\cal T}} \left( p_{ij}({\bf V}_{ij} | {\bf V}_i) \right)^{\mathbb{I}(s^i_{t},s^j_{t+1})} \right) 1^{\mathbb{I}(s_{t}^*)}\\
p_1({\bf V}_1) &
=\!\!\! \prod_{V \in {\bf V}_1}\!\!\! p(V | \pa_{{\cal G}_1}(V)); \hspace{0.05cm}
p_{ij}({\bf V}_{ij} | {\bf V}_i) 
=\!\!\! \prod_{V \in {\bf V}_{ij}} \!\!\!\!\! p(V | \pa_{{\cal G}_{ij}}(V))
\end{align*}
}%

An intervention in a PDSEM is defined on a set of treatment variables ${\bf A} \equiv \bigcup_{(i,j) \in {\cal T}} {\bf A}_{ij}$ and a set to values ${\bf a}$ with the property that for any
$(i,j),(k,j) \in {\cal T}$,
the same values ${\bf a}_j$ are being set to ${\bf A}_{ij}$ and ${\bf A}_{ij}$.
Define ${\bf Y}_{ij}$ in each transition graph ${\cal G}_{ij}$ to be all variables in that state not in ${\bf A}_{ij}$, with their corresponding values being ${\bf y}_j$, their union being ${\bf Y}$, and the values of the union being ${\bf y}$.

A new counterfactual distribution $p_{\infty}({\bf Y}({\bf a}))$ is obtained from the counterfactual initial state distribution $p_1({\bf Y}_1({\bf a}_1))$,
and transition distributions $p_{ij}({\bf Y}_{ij}({\bf a}_{j}) | {\bf Y}_{i}({\bf a}_i))$ %
as:

\vspace{-0.5cm}
{\small
	\setlength{\abovedisplayskip}{5pt} \setlength{\belowdisplayskip}{5pt}
\begin{align}
p_1({\bf Y}_1({\bf a}_1)) \prod_{t=1}^{\infty} \left( \prod_{(i,j) \in {\cal T}} \left( p_{ij}({\bf Y}_{ij}({\bf a}_{j}) | {\bf Y}_i({\bf a}_i)) \right)^{\mathbb{I}(s^i_{t},s^j_{t+1})} \right)
\!\!
1^{\mathbb{I}(s_{t}^*)}
\label{eqn:cntfl-pdsem}
\end{align}
}%
\vspace{-0.5cm}

Individual counterfactual distributions are obtained using standard structural equation replacement semantics.

Since the initial state and transitions are defined using structural equations, we obtain the following identification result, which generalizes the DBN g-formula (\ref{eqn:g-dbn}) to PDSEMs.
\begin{lem}
\label{eqn:pdsem-g}
Given a fully observed PDSEM, each factor of the distribution $p_{\infty}({\bf Y}({\bf a}))$ is identified from $p_{\infty}({\bf V})$ as:
{\small
\begin{align}
\notag
p_1({\bf Y}_1({\bf a}_1)) &\equiv \prod_{V \in {\bf Y}_1 \setminus {\bf A}_1} p_{1}(V | \pa_{{\cal G}_1}(V)) \Big\vert_{{\bf A}_1 = {\bf a}_1} \\
p_{ij}({\bf Y}_{ij}({\bf a}_{j}) | {\bf Y}_i({\bf a}_i)) &\equiv \prod_{V \in {\bf Y}_{ij} \setminus {\bf A}_{j}} p_{ij}(V | \pa_{{\cal G}_{ij}}(V)) \Big\vert_{\substack{{\bf A}_i = {\bf a}_i, \\{\bf A}_j = {\bf a}_j}}
\end{align}
}
\end{lem}

\section{PDSEMs With Hidden Variables} \label{pdsem-admgs}
 In extending causal inference to latent variable PDSEMs, in addition to Assumption \ref{assump:fixed-observed} and Assumption \ref{assump:state-vars}, we also assume
the probabilities of any state transition trajectories are observed.
\begin{assump} \label{assump:transition-observed}
	The variables $S_{ij}$ for any $(i,j) \in {\cal T}$ governing state transition probabilities are observed.
\end{assump}
The latent variable PDSEMs then decompose into an initial state and a set of transitions such that causal inference results may be stated without loss of generality using latent projection ADMGs (and CADMGs) of appropriate DAGs and CDAGs.
In addition, the fact that variables $S_{ij}$ are observed implies we can evaluate counterfactual state transition probabilities, provided they are identified.

Formally, fix a PDSEM defined given the initial state DAG is ${\cal G}$ on ${\bf V}_1, {\bf H}_1$ and the set of transition CDAGs ${\cal G}_{ij}$ on ${\bf V}_{ij},{\bf H}_{ij}$ given ${\bf V}_i$, for all $(i,j) \in {\cal T}$, such that $S_1 \in {\bf V}_1$, $S_{ij} \in {\bf V}_{ij}$ for every $(i,j) \in {\cal T}$,
for every $j$ and all $(i,j),(k,j) \in {\cal T}$, ${\bf H}_{ij} = {\bf H}_{kj}$ and ${\bf V}_{ij} = {\bf V}_{kj}$.
We assume the variables
${\bf V} \equiv \{ {\bf V}_1 \} \cup
\bigcup_{(i,j) \in {\cal T}} {\bf V}_{ij}$,
and
${\bf H} \equiv \{ {\bf H}_1 \} \cup 
\bigcup_{(i,j) \in {\cal T}} {\bf H}_{ij}$ 
are observed, and hidden, respectively.

Given this definition of a latent variable PDSEM, the observed data distribution $p_{\infty}({\bf V})$ is obtained from applying the usual transition probabilities to the margin at the initial state $p_1({\bf V}_1) \equiv \sum_{{\bf H}_1} p_1({\bf V}_1 \dot{\cup} {\bf H}_1)$, and the margins of all transition probabilities $p_{ij}({\bf V}_{ij} | {\bf V}_i) \equiv \sum_{{\bf H}_{ij}} p_{ij}({\bf V}_{ij} \dot{\cup} {\bf H}_{ij} | {\bf V}_i)$.

As before, fix a set of observed treatment variables ${\bf A}$, which is the union of $\{ {\bf A}_{ij} : (i,j) \in {\cal T} \}$, such that the same values ${\bf a}_j$ are set to ${\bf A}_{ij}, {\bf A}_{kj}$ for any $(i,j),(k,j) \in {\cal T}$, and the set of outcomes ${\bf Y}_{ij} = {\bf V}_{ij} \setminus {\bf A}_{ij}$ for any $(i,j) \in {\cal T}$, with ${\bf Y}$ the union of $\{ {\bf Y}_{ij} : (i,j) \in {\cal T} \}$.

Given the way the latent variable PDSEM was defined, identification theory for $p_{\infty}({\bf Y}({\bf a}))$ reduces to identification theory for $p_1({\bf Y}_1({\bf a}_1))$ in the latent projection ADMG ${\cal G}_1$ on ${\bf V}_1$, and $p_{ij}({\bf Y}_{ij}({\bf a}_{j}) | {\bf V}_i({\bf a}_i))$ in the latent projection CADMG ${\cal G}_{ij}$ on ${\bf V}_{ij}$ given ${\bf V}_i$, as follows:

\begin{lem} \label{eqn:hidden-pdsem-id}
Under Assumptions \ref{assump:fixed-observed}, \ref{assump:state-vars} and \ref{assump:transition-observed}, given a latent variable PDSEM represented by ${\cal G}_1$ and $\{ {\cal G}_{ij} : (i,j) \in {\cal T} \}$,
$p_{\infty}({\bf Y}({\bf a}))$ is identified from $p_{\infty}({\bf V})$ if and only if every bidirected component in ${\cal G}_{1{\bf Y}_1}$ is intrinsic in ${\cal G}_1$, and
every bidirected component in ${\cal G}_{ij{\bf Y}_j}$ is intrinsic in ${\cal G}_{ij}$ for every $i$ and $j$.  Moreover, if $p_{\infty}({\bf Y}({\bf a}))$ is identified, it is equal to
{\small
\begin{gather}
p_1({\bf Y}_1({\bf a}_1)) \prod_{t=1}^{\infty} \left( \prod_{(i,j) \in {\cal T}} \left( p_{ij}({\bf Y}_{ij}({\bf a}_{j}) | {\bf Y}_i({\bf a}_i)) \right)^{\mathbb{I}(s^i_{t-1},s^j_{t})} \right)
\!\!
1^{\mathbb{I}(s_{t-1}^*)}
\label{eqn:admg-pdsem-top}
\shortintertext{where}
p_1({\bf Y}_1({\bf a}_1)) = \prod_{{\bf D} \in {\cal D}({\cal G}_{1{\bf Y}^*_1})} q^1_{\bf D}({\bf D} | \pa^s_{{\cal G}_1}({\bf D})) \Big\vert_{{\bf A}_1 = {\bf a}_1},
\label{eqn:admg-pdsem-prior}
\end{gather}}%
where each kernel $q^1_{\bf D}({\bf D} | \pa^s_{{\cal G}_1}({\bf D}))$ is in the nested Markov factorization of $p_1({\bf V}_1)$ with respect to ${\cal G}_1$, and
{\small
\begin{align}
p_{ij}({\bf Y}_{ij}({\bf a}_{j}) | {\bf Y}_i({\bf a}_i)) = \quad \prod_{\mathclap{{\bf D} \in {\cal D}({\cal G}_{{\bf V}_{ij} \setminus {\bf A}_{ij}})}} \quad q^{ij}_{\bf D}({\bf D} | \pa^s_{{\cal G}_{ij}}({\bf D})) \Big\vert_{\substack{{\bf A}_i={\bf a}_i, \\ {\bf A}_j={\bf a}_j}}
\label{eqn:admg-pdsem-transition}
\end{align}
}
where each kernel $q^{ij}_{\bf D}({\bf D} | \pa^s_{{\cal G}_{ij}}({\bf D}))$ is in the nested Markov factorization of $p_{ij}({\bf V}_{ij} | {\bf V}_i)$ with respect to 
${\cal G}_{ij}$.
\label{eqn:pdsem-admg-id}
\end{lem}
An example of a hidden variable PDSEM and identifying functionals are given in Section \ref{nested-markov} of the Appendix.

\section{Experiments} \label{experiments}
\vspace*{-2mm}
\subsection{Simulation of a latent variable PDSEM}\label{exp-sim}
\vspace*{-1mm}
We show how statistical inference may be performed in the example presented in Section~\ref{pdsem-dags-toy} and Fig.~\ref{fig:dags-sim}, altered to include latent variables. The system has 
states $\{s^1,s^2,s^3\}$ and three variables in each state $\{A,B,C\}$. Additionally, $s^2$ has a hidden common cause of $A$ and $B$. This is represented by the red (dotted) bidirected edge $A \leftrightarrow B$ in the latent projected ADMG shown in Fig.~\ref{fig:dags-sim}(c). Patient health status $A$, surgeon experience $B$, 
and duration of the stage of surgery $C$, are all continuous variables. State and transition graphs are identical to those in Fig.~\ref{fig:dags-sim}.

Two sets of parameters associated with a generative model of this kind are $p(S_{t+1} = s^j | S_t = s^i, {\bf V}_t)$, where $s^i_{t} \rightarrow s^j_{t+1}$ is a transition allowed by the model and $p(V^{ij}_{t+1} = v | S_{t+1}=s^j, S_t = s^i, {\bf V}_t)$, where $V^{ij} \in \{A^{ij},B^{ij},C^{ij}\}$, 
and, $s^i_{t} \rightarrow s^j_{t+1}$ is an allowed transition. These parameters 
are chosen to be reasonable for the surgery application, yielding a distribution Markov relative to appropriate graphs.

A dataset of $N = 10000$ ``surgeries'' was simulated, all with the same initial state. Transition probabilities were generated using a logistic regression on variables in the current state, with transitions eventually
terminating at the absorbing state.
Each variable $V_i$ is generated from a set of linear structural equations with correlated errors. 
Using generated data, state transition probabilities were estimated using maximum likelihood. 
Parameters for the structural equation model were estimated using the RICF algorithm \citep{drton2009computing}, implemented in the Ananke \footnote{https://ananke.readthedocs.io/en/latest/index.html} package \citep{ananke}.

We used the PDSEM to consider the causal impact of surgeon experience (measured by total operating time in their career) on average surgery length. This outcome is easy to measure, and is known to serve as an informative proxy for other measures of surgery quality, such as follow-up assessments of quality of life \citep{rambachan2013impact,jackson2011does}.
We assessed this causal question by generating two sets of sampled surgery trajectories where, in each stage of the surgery, the surgeon was intervened to have higher (vs. lower) career operating time by one unit. These trajectories may be viewed as a Monte Carlo sampling scheme for evaluating the functional given by (\ref{eqn:admg-pdsem-top}), (\ref{eqn:admg-pdsem-prior}) and (\ref{eqn:admg-pdsem-transition}).  The comparison of these two sets of trajectories may be viewed as a generalization of the \emph{average causal effect (ACE)} from classical longitudinal causal models to PDSEMs.

The results are shown in Fig. ~\ref{fig:sim-intervention-plot}. Surgeries performed by experienced surgeons are shorter ( $\mu = 5.79$, $\mathbf{q}_{0.05} = 3, \mathbf{q}_{0.95} = 13$) than those performed by trainees ($\mu = 7.02$, $\mathbf{q}_{0.05} = 3$, $\mathbf{q}_{0.95} = 17$) where $\mathbf{q}_{p}$ denotes the $p^{\textrm{th}}$ quantile. Surgeries performed by trainees have higher variance.

\begin{figure}{}
	\begin{center}
		\centerline{\includegraphics[width=\columnwidth]{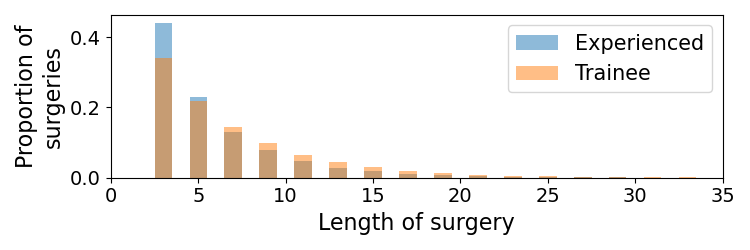}}
		\caption{
Histograms of the number of transitions in a surgery under two different interventions: when a more experienced surgeon performs the entire procedure, and when a less experienced trainee performs the entire procedure.
}
		\label{fig:sim-intervention-plot}
	\end{center}
	\vspace{-1.7em}
\end{figure}

\subsection{Data Application of the PDSEM} \label{exp-data}
\vspace*{-1mm}
We now illustrate how a PDSEM may be applied to analyze data obtained from a surgery. Our dataset consists of 236 septoplasty procedures conducted at our institution's research hospital. A total of 57343 timestamped records were collected, including tool and personnel activity. Surgeries consist of six distinct phases: $s^1$ (opening of the septum), $s^2$ (raising septal flaps), $s^3$ (removal of deviated septal cartilage and bone), $s^4$ (reconstruction), $s^5$ (closing of the incision), and $s^6$ (other activity). An artificial absorbing state $s^{\text{end}}$ represents the end of procedures. Procedures are often led by an attending, with a surgeon trainee assisting. Of the surgeries, 42.79\% of them were performed fully by the leading attending; the others by a team. Also, attending surgeons perform for 64.98\% of all operating time and trainees the rest. Twelve different surgical tools were tracked for use. Each phase of the surgery requires different techniques and tools. The progression of the surgery is not monotonic -- surgeons commonly revisit earlier stages.
The state transition diagram representing allowed state transitions is presented in Fig \ref{fig:transition-diagram-data}.
We chose to discretize all variables into two categories. Model parameters were estimated by maximum likelihood. More details about the data and model can be found in Section \ref{septoplasty} of the Appendix.


\begin{figure}[h]
	\begin{center}
		\centerline{\includegraphics[width=\columnwidth]{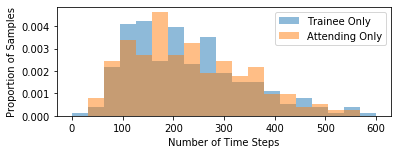}}
		\caption{Histograms of hypothetical surgeries performed only by a junior trainee surgeon (blue) versus hypothetical surgeries performed only by a senior attending surgeon (orange). Surgeries performed by the attending are slightly longer ($\mu= 244.3.91, \sigma=139.9$) than those of the trainee ($\mu = 233.5, \sigma=125.9$).}
		\label{fig:intervention}
	\end{center}

	\begin{center}
		\begin{tikzpicture}[>=stealth, scale=.5]
		\tikzstyle{format} = [draw=none, very thick, circle, minimum size=3mm,
		inner sep=0pt]		
		
		\begin{scope}[xshift=3cm,yshift=-2.5cm]
		\path[->, thick]
		node[format] (s1) {$s^{1}$}
		node[format, above of=s1,xshift=0cm,yshift=-0.0cm] (s2) {$s^{2}$}
		node[format, right of=s1,xshift=0.3cm,yshift=-0.0cm] (s3) {$s^{3}$}
		node[format, right of=s2,xshift=0.3cm,yshift=-0.0cm] (s4) {$s^{4}$}
		
		node[format, right of=s3,xshift=0cm,yshift=-0.0cm] (s5) {$s^{5}$}
		node[format, right of=s4,xshift=1.2cm,yshift=-0.0cm] (s6) {$s^{6}$}
		node[format, right of=s5,xshift=0cm,yshift=-0.0cm] (s7) {$s^{\textrm{end}}$}
		
		(s1) edge[black] (s2)
		(s2) edge[black] (s3)
		(s3) edge[black] (s4)
		(s3) edge[black] (s6)
		(s4) edge[black, bend left = 10] (s5)
		(s4) edge[black, bend left = 30] (s6)
		(s4) edge[black, bend left = 10] (s7)
		(s5) edge[black, bend left=20] (s4)
		(s5) edge[black] (s6)
		(s5) edge[black] (s7)
		(s6) edge[black, bend left = 0] (s4)
		(s1) edge[loop, out=230,in=300,looseness=5] (s1)
		(s3) edge[loop, out=230,in=300,looseness=5] (s3)
		(s5) edge[loop, out=230,in=300,looseness=5] (s5)
		(s2) edge[loop, in=60,out=130,looseness=5] (s2)
		(s4) edge[loop, in=60,out=130,looseness=5] (s4)
		(s6) edge[loop, in=60,out=130,looseness=5] (s6)
		
		;
		\end{scope}
		\end{tikzpicture}
	\end{center}
	\caption{The state transition diagram for the surgery data application.}
	\label{fig:transition-diagram-data}
\end{figure}

As before, we considered the causal impact of surgeon experience on average length of surgery, evaluated by considering counterfactual trajectories and comparing to trajectories observed in the data. Estimation of  $p(s_t | s_{t-1}, \mathbf{v}_{t-1})$ at all levels of $s_{t-1}, \mathbf{v}_{t-1}$ is not always possible due to finite sample limitations. To address this, we apply additive smoothing to $p(s_t | s_{t-1}, \mathbf{v}_{t-1})$, based on the empirical distribution $p(s_t | s_{t-1})$.
Goodness of fit is illustrated in Fig \ref{fig:real-vs-model} of the Appendix and results are presented in Fig \ref{fig:intervention}. We have made considerable assumptions in modeling our PDSEM and have closely matched the generative model to the empirical distribution (Fig \ref{fig:real-vs-model}). We observe that the causal effect of surgeon skill on surgery length, given our learned parameters, 
is close to zero.  This indicates that policies that govern the trade-off between the need to train surgeons, and overall surgery quality (as quantified by our 
outcome) are effective at our institution.

Generalizing statistical inference in PDSEMs with hidden variables to likelihoods based on parameterizations of the nested Markov model \citep{richardson17nested} presents a number of open problems, especially for variables that are not binary or not multivariate normal. We discuss these issues in Section \ref{septoplasty} of the Appendix.


\section{Conclusions} \label{conclusions}
\vspace{-0.15cm}
In this paper, we have introduced the Path Dependent Structural Equation Model (PDSEM) for longitudinal data which unifies complex state structure from DBNs and complex state transition dynamics from MDPs. It can also be seen as a graphical model generalizing the dynamics of a Markov chain with state-specific dynamics. We have described counterfactuals associated with these causal models that can alter the subsequent temporal evolution of the system, identification theory for such counterfactuals in terms of the observed data distribution, and described estimation.
We showed the utility of the model in clinical settings using simulations as well as real data from a septoplasty surgical procedure.
Developing novel methods for efficient Monte Carlo sampling based statistical inference for hidden variable versions of PDSEMs based on the nested Markov model is a promising area of future work.

\section*{Acknowledgements}
The data on nasal septoplasty used in this paper were collected within a study supported by research grants from the National Institutes of Health(NIH; R21-DE022656 and R01-DE05265; PI: Dr. Masaru Ishii, MD, PhD). We thank Dr. Greg Hager, Dr. Masaru Ishii, Dr. Swaroop Vedula and the rest of their team at Johns Hopkins University for access to this data and their valuable insights on surgical training.

\clearpage

\bibliography{pdcm}

\appendix
\onecolumn
{\huge\bfseries Appendix}
\section{Graph preliminaries}\label{graph-prelims}
Let capital letters $X$ denote random variables, and let lower case letters $x$ values of X. Sets of random variables are denoted ${\bf V}$, and sets of values ${\bf v}$.
For a subset ${\bf A} \subseteq {\bf V}$, ${\bf v}_{{\bf A}}$ denotes the subset of values in ${\bf v}$ of variables in ${\bf A}$.
Domains of $X$ and ${\bf X}$ are denoted by ${\mathfrak X}_X$ and ${\mathfrak X}_{{\bf X}}$, respectively.

Standard genealogic relations on graphs are as follows: parents, children, descendants, siblings and ancestors of $X$ in a graph ${\cal G}$ are denoted by
$\pa_{\cal G}(X), \ch_{\cal G}(X), \de_{\cal G}(X), \si_{\cal G}(X), \an_{\cal G}(X)$, respectively \citep{lauritzen96graphical}.  These relations are defined disjunctively for sets, e.g.
$\pa_{\cal G}({\bf X}) \equiv \bigcup_{X \in {\bf X}} \pa_{\cal G}(X)$.  By convention, for any $X$, $\an_{\cal G}(X) \cap \de_{\cal G}(X) \cap \dis_{\cal G}(X) = \{ X \}$.

We will also define the set of \emph{strict parents} as follows: $\pas_{\cal G}({\bf X}) = \pa_{\cal G} ({\bf X}) \setminus {\bf X}$.
Given any vertex $V$ in an ADMG ${\cal G}$, define the \emph{ordered Markov blanket} of $V$ as $\mb_{\cal G}(V) \equiv (\dis_{\cal G}(V) \cup \pa_{\cal G}(\dis_{\cal G}(V)) ) \setminus V$.
Given a graph ${\cal G}$ with vertex set ${\bf V}$, and ${\bf S} \subseteq {\bf V}$, define the \emph{induced subgraph} $\mathcal{G}_{{\bf S}}$ to be a graph containing the vertex set ${\bf S}$ and all edges in ${\cal G}$ among elements in ${\bf S}$.

In the subsequent discussion, we will denote an ADMG ${\cal G}$ on ${\bf V}$ by notation ${\cal G}({\bf V})$, and a CADMG ${\cal G}$ on ${\bf V}$ given ${\bf W}$ by notation ${\cal G}({\bf V},{\bf W})$.

\section{The Nested Markov Factorization}\label{nested-markov}
It is recommended that the reader look up notation for graphs in Section \ref{graph-prelims} of the Appendix to follow this section.

\subsection{Why do we need an alternative factorization?}
A hidden variable CDAG ${\cal G}({\bf V} \cup {\bf H}, {\bf W})$ may be used to define a factorization on distributions $p({\bf V} | {\bf W})$ in terms of the CDAG as: $p({\bf V}|{\bf W}) = \sum_{{\bf H}} \prod_{V \in {\bf V} \cup {\bf H}} p(V | \pa_{\cal G}(V))$. However, inferences may be sensitive to assumptions made about the state spaces for the unobserved variables and the latent variable model may contain singularities at which asymptotics are irregular \citep{drton2009likelihood}. Additionally, such a model does not form a tractable search space: an arbitrary number of hidden variables and associated structures may be incorporated that are consistent with observed data distributions.

Alternatively, a factorization of the marginal distribution $p({\bf V}|{\bf W})$ can be defined directly on the latent projection CADMG ${\cal G}({\bf V},{\bf W})$. This \emph{nested Markov factorization}, described in \citep{richardson17nested} completely avoids modeling hidden variables, and leads to a regular likelihood in special cases \citep{evans18smooth}. It captures all equality constraints a hidden variable CDAG factorization imposes on the observed margin $p({\bf V}|{\bf W})$ \citep{shpitser18sem}.  In addition, $p({Y}({a})|{\bf W})$ (an interventional distribution given a fixed context ${\bf W}$) identified in a hidden variable causal model represented by ${\cal G}({\bf V} \cup {\bf H}, {\bf W})$ is always equal to a modified version of a nested factorization \citep{richardson17nested} associated with ${\cal G}({\bf V},{\bf W})$, described here. 

\subsection{The nested Markov factorization}
The nested Markov factorization of $p({\bf V}|{\bf W})$ with respect to a CADMG ${\cal G}({\bf V},{\bf W})$ links \textit{kernels}, mappings derived from $p({\bf V}|{\bf W})$ and CADMGs derived from ${\cal G}({\bf V},{\bf W})$ via a \textit{fixing} operation.\\

{\bf Kernel:} A kernel $q_{\bf V}({\bf V} | {\bf W})$ is a mapping from values in ${\bf W}$ to normalized densities over ${\bf V}$ \citep{lauritzen96graphical}. A conditional distribution is a familiar example of a kernel, in that $\sum_{{\bf v} \in {\bf V}} q_{\bf V}({\bf v} | {\bf w}) = 1$. Conditioning and marginalization are defined in kernels in the usual way: For ${\bf A} \subseteq {\bf V}$, $q_{\bf V}({\bf A} | {\bf W}) \equiv \sum_{{\bf V} \setminus {\bf A}} q_{\bf V}({\bf V} | {\bf W})$ and $q_{\bf V}({\bf V} \setminus {\bf A} | {\bf A} \cup {\bf W}) \!\equiv\! \frac{q_{\bf V}({\bf V} | {\bf W})}{q_{\bf V}({\bf A} | {\bf W})}$. \\

{\bf Fixability and the fixing operator:} A variable $V \in {\bf V}$ in a CADMG ${\G}$ is fixable if $\text{de}_{\G}(V) \cap \text{des}_{\G}(V) = \emptyset$. In other words, V is fixable if paths $V \leftrightarrow ... \leftrightarrow B$ and $V \to ... \to B$ do not both exist in ${\G}$ for any $B \in {\bf V} \textbackslash \{V \}$.
 
We define a fixing operator $\phi_V({\cal G})$ for graphs, and a fixing operator $\phi_V(q; {\cal G})$ for kernels. Given a CADMG ${\cal G}({\bf V},{\bf W})$, with a fixable $V \in {\bf V}$, $\phi_V({\cal G}({\bf V},{\bf W}))$ yields a new CADMG ${\cal G}({\bf V} \setminus \{ V \}, {\bf W} \cup \{ V \})$ obtained from ${\cal G}({\bf V}, {\bf W})$ by moving $V$ from ${\bf V}$ to ${\bf W}$, and removing all edges with arrowheads into $V$.  Given a kernel $q_{{\bf V}}({\bf V} | {\bf W})$, and a CADMG ${\cal G}({\bf V}, {\bf W})$, the operator
$\phi_V(q_{{\bf V}}({\bf V} | {\bf W}), {\cal G}({\bf V}, {\bf W}))$ yields a new kernel:
$$q_{{\bf V} \setminus \{ V \}}({\bf V} \setminus \{ V \} | {\bf W} \cup \{ V \}) \equiv
\frac{
	q_{{\bf V}}({\bf V} | {\bf W})
}{
q_{{\bf V}}(V | \mb_{\cal G}(V))
}$$

{\bf Fixing sequences:} A sequence $\langle V_1, \ldots, V_k \rangle$ is said to be \emph{valid} in ${\cal G}({\bf V},{\bf W})$ if $V_1$ fixable in ${\cal G}({\bf V}, {\bf W})$, $V_2$ is fixable in $\phi_{V_1}({\cal G}({\bf V}, {\bf W}))$, and so on.  If any two sequences $\sigma_1, \sigma_2$ for the same set ${\bf S} \subseteq {\bf V}$ are fixable in ${\cal G}$, they lead to the same CADMG. The graph fixing operator can be extended to a set ${\bf S}$: $\phi_{{\bf S}}({\cal G})$. This operator is defined as applying the vertex fixing operation in any valid sequence $\sigma$ for set ${\bf S}$.

Given a sequence $\sigma_{{\bf S}}$, define $\eta(\sigma_{{\bf S}})$ to be the first element in $\sigma_{{\bf S}}$, and $\tau(\sigma_{{\bf S}})$ to be the subsequence of $\sigma_{{\bf S}}$ containing all elements but the first. Given a sequence $\sigma_{{\bf S}}$ on elements in ${\bf S}$ valid in ${\cal G}({\bf V}, {\bf W})$, the kernel fixing operator $\phi_{\sigma_{{\bf S}}}(q_{{\bf V}}({\bf V} | {\bf W}), {\cal G}({\bf V}, {\bf W}))$ is defined to be equal to $q_{{\bf V}}({\bf V} | {\bf W})$ if $\sigma_{{\bf S}}$ is the empty sequence, and
$\phi_{\tau(\sigma_{{\bf S}})}( \phi_{\eta(\sigma_{{\bf S}})}(q_{{\bf V}}({\bf V} | {\bf W}); {\cal G}({\bf V}, {\bf W})), \phi_{\eta(\sigma_{{\bf S}})}({\cal G}({\bf V}, {\bf W})))$ otherwise.\\

{\bf Reachability:} Given a CADMG ${\cal G}({\bf V}, {\bf W})$, a set ${\bf R} \subseteq {\bf V}$ is called \emph{reachable} if there exists a sequence for ${\bf V} \setminus {\bf R}$ valid in ${\cal G}({\bf V},{\bf W})$. In other words, if ${\bf S}$ is fixable in $\G$, ${\bf V} \setminus {\bf S}$ is reachable. \\

{\bf Intrinsic sets:} A set ${\bf R}$ reachable in ${\cal G}({\bf V},{\bf W})$ is \emph{intrinsic} in ${\cal G}({\bf V},{\bf W})$ if $\phi_{{\bf V} \setminus {\bf R}}({\cal G})$ contains a single district, ${\bf R}$ itself.  The set of intrinsic sets in a CADMG ${\cal G}$ is denoted by ${\cal I}({\cal G})$.\\

{\bf Nested Markov factorization:} A distribution $p({\bf V}|{\bf W})$ is said to obey the \emph{nested Markov factorization} with respect to the CADMG ${\cal G}({\bf V},{\bf W})$ if there exists a set of kernels of the form $\{ q_{{\bf S}}({\bf S} | \pa_{\cal G}({\bf S}) ) : {\bf S} \in {\cal I}({\cal G}) \} \}$ such that for every valid sequence $\sigma_{{\bf R}}$ for a reachable set ${\bf R}$ in ${\cal G}$, we have:
\begin{align*}
\phi_{\sigma_{{\bf R}}}(p({\bf V}|{\bf W}); {\cal G}({\bf V},{\bf W})) 
= \prod_{{\bf D} \in {\cal D}(\phi_{{\bf R}}({\cal G}({\bf V},{\bf W})))} q_{{\bf D}}({\bf D} | \pas_{\cal G}({\bf D}) )
\end{align*}

If a distribution obeys this factorization, then for any reachable ${\bf R}$, any two valid sequences on ${\bf R}$ applied to $p({\bf V}|{\bf W})$ yield the same kernel $q_{{\bf R}}({\bf R} | {\bf V} \setminus {\bf R})$.  Hence, kernel fixing may be defined on sets, just as graph fixing.  In this case, for every
${\bf D} \in {\cal I}({\cal G})$, $q_{{\bf D}}({\bf D} | \pas_{\cal G}({\bf D}) ) \equiv \phi_{{\bf V} \setminus {\bf D}}(p({\bf V}|{\bf W}); {\cal G}({\bf V},{\bf W}))$.

The \emph{district factorization} or \emph{Tian factorization} of $p({\bf V}|{\bf W})$ results from the nested factorization:
\begin{align*}
p({\bf V}|{\bf W})
&= \prod_{{\bf D} \in {\cal D}({\cal G}({\bf V},{\bf W}))} q_{{\bf D}}({\bf D} | \pas_{\cal G}({\bf D})) \\
&
= \prod_{{\bf D} \in {\cal D}({\cal G}({\bf V},{\bf W}))} \left( \prod_{D \in {\bf D}} p(D \mid \pre_{\prec}(D)) \right),
\end{align*}
where $\pre_{\prec}(D)$ is the set of predecessors of $D$ according to a topological total ordering $\prec$.  Each factor $\prod_{D \in {\bf D}} p(D \mid \pre_{\prec}(D))$ is only a function of ${\bf D} \cup \pa_{\cal G}({\bf D})$ under the nested factorization.

An important result in \citep{richardson17nested} states that if $p({\bf V} \cup {\bf H}|{\bf W})$ obeys the factorization for a CDAG $\G({\bf V} \cup {\bf H},{\bf W})$, then $p({\bf V}|{\bf W})$ obeys the nested factorization for the latent projection CADMG ${\cal G}({\bf V},{\bf W})$.

\subsection{Identification}
Not every interventional distribution $p({\bf Y}({\bf a}))$ is identified in a hidden variable causal model.
However, \emph{every} $p({\bf Y}({\bf a})|{\bf W})$ identified from $p({\bf V}|{\bf W})$ can be expressed as a modified nested factorization as follows:
\begin{align*}
p({\bf Y}({\bf a})|{\bf W})
&=\!\!
\sum_{{\bf Y}^* \setminus {\bf Y}} \prod_{{\bf D} \in {\cal D}(
	{\cal G}_{{\bf Y}^*}
	)} p({\bf D} | \doo(\pas_{\cal G}({\bf D}) )) \vert_{{\bf A} = {\bf a}}\\
&=\!\!
\sum_{{\bf Y}^* \setminus {\bf Y}} \prod_{{\bf D} \in {\cal D}(
	{\cal G}_{{\bf Y}^*}
	)} \phi_{{\bf V} \setminus {\bf D}}(p({\bf V}|{\bf W}); {\cal G}({\bf V},{\bf W})) \vert_{{\bf A} = {\bf a}}
\end{align*}
where ${\bf Y}^* \equiv \an_{{\cal G}({\bf V}({\bf a}), {\bf W})}({\bf Y}) \setminus {\bf a}$. That is, $p({\bf Y}({\bf a})|{\bf W})$ is only identified if it can be expressed as a factorization, where every
piece corresponds to a kernel associated with a set intrinsic in $\G({\bf V},{\bf W})$. Moreover, no piece in this factorization contains elements of $\bf A$ as random variables.

\subsection{Example of the nested factorization of a hidden variable PDSEM}\label{sec:appendix-example}

A hidden variable PDSEM can be unrolled into a latent-projected ADMG if the model obeys restrictions given in Section \ref{pdsem-admgs}. For instance, Fig. \ref{fig:admg-pdsem-example} in this Appendix shows an example where the first two states of the system involve hidden variables. In particular, the system at $s^2$ is the \textit{front-door-graph} previously encountered in Section \ref{bkgd}. Transition graphs are in Fig. \ref{fig:admg-pdsem-example}(c)-(e). 

The nested factorization for the initial graph in Fig.~\ref{fig:admg-pdsem-example} (a) has intrinsic sets
\begin{align*}
(a)&: \{ A_1 \}, \{ B_1 \}, \{ C_1 \}, \{ A_1, B_1 \}, \{ S_1 \}
\end{align*}
with corresponding kernels
\begin{align}
(a)&: q_{A_1}(A_1) \equiv p(A_1); q_{B_1}(B_1) = p(B_1); q_{C_1}(C_1 | A_1, B_1) \equiv p(C_1 | A_1, B_1); q_{A_1,B_1}(A_1,B_1) \equiv p(A_1,B_1); q_{S_1}(S_1) \equiv p(S_1).
\label{eqn:prior}
\end{align}
Similarly, the nested factorizations for the transition graphs in Fig.~\ref{fig:admg-pdsem-example} (c),(d),(e) have intrinsic sets:
\begin{align*}
(c)&: \{ A_{12} \}, \{ B_{12} \}, \{ C_{12} \}, \{ A_{12}, C_{12} \}, \{ S_{12} \}\\
(d)&: \{ A_{23} \}, \{ B_{23} \}, \{ C_{23} \}, \{ S_{23} \}\\
(e)&: \{ A_{21} \}, \{ B_{21} \}, \{ A_{21}, B_{21} \}, \{ C_{21} \}, \{ S_{21} \},
\end{align*}
with corresponding kernels
\begin{align}
\notag
(c)&: q_{A_{12}}(A_{12} | C_1) \equiv p(A_{12} | C_1); q_{B_{12}}(B_{12} | A_{12})) \equiv p(B_{12} | A_{12}); q_{C_{12}}(C_{12} | C_1, B_{12}) \equiv
	\sum_{A_{12}} p(C_{12} | B_{12},A_{12},C_1) p(A_{12} | C_1);\\
\notag
	& \hspace{0.3cm} q_{A_{12},C_{12}}(A_{12},C_{12} | B_{12},C_1) \equiv p(C_{12} | B_{12},A_{12},C_1) p(A_{12} | C_1);
		q_{S_{12}}(S_{12} | C_{12}) \equiv p(S_{12} | C_{12}).
	\\
\notag
(d)&: q_{A_{23}}(A_{23} | A_2, C_2) \equiv p(A_{23} | A_2, C_2); q_{B_{23}}(B_{23} | B_2, A_{23}) \equiv p(B_{23} | B_2, A_{23}); q_{C_{23}}(C_{23} | C_2, A_{23}) \equiv
		p(C_{23} | C_2, A_{23});\\
\notag
	& \hspace{0.3cm} q_{S_{23}}(S_{23}) \equiv p(S_{23}).
	\\
\notag
(e)&: q_{A_{21}}(A_{21} | A_2) \equiv p(A_{21} | A_2); q_{B_{21}}(B_{21} | B_2) \equiv p(B_{21} | B_2); q_{A_{21},B_{21}}(A_{21},B_{21}) \equiv p(A_{21}, B_{21});\\
	& \hspace{0.3cm} q_{C_{21}}(C_{21} | C_2, B_{21}, A_{21}) \equiv p(C_{21} | C_2, B_{21}, A_{21}); q_{S_{21}}(S_{21}) \equiv p(S_{21}).
\label{eqn:transition}
\end{align}

Applying the Nested Markov factorization on the trajectory in \ref{fig:admg-pdsem-example} (f), we obtain the following factorization:

\vspace*{-3mm}
{\small 
	\setlength{\abovedisplayskip}{5pt} \setlength{\belowdisplayskip}{5pt}
	\begin{align*}
&	p(A_1, B_1, C_1) \cdot p(A_{12},B_{12},C_{12} | A_1, B_1, C_1) \cdot p(A_{23}, B_{23}, C_{23} | A_{12}, B_{12}, C_{12}) \\
=&
\underbrace{\left\{
q_{A_1,B_1}(A_1,B_1) q_{C_1}( C_1\mid A_1,B_1) 
\right\}}_{(a)}
 \cdot
\underbrace{\left\{
q_{A_{12},C_{12}}(A_{12},C_{12} | B_{12},A_1,C_1)
q_{B_{12}}(B_{12} | A_{12})
\right\}}_{(c)} \cdot\\
&
\underbrace{\left\{
q_{A_{23}}(A_{23}| A_{12},C_{12}) \cdot q_{A_{23}}(B_{23}| B_{12},A_{23}) \cdot q_{C_{23}}(C_{23}| C_{12},A_{23})
\right\}}_{(d)},
	\end{align*}
}
where the kernels are given in (\ref{eqn:prior}) and (\ref{eqn:transition}) above.

\begin{figure}[t]
	\begin{center}
		\begin{tikzpicture}[>=stealth, node distance=1.4cm]
		\tikzstyle{format} = [draw=none, very thick, circle, minimum size=6mm,
		inner sep=0pt]
		\tikzstyle{square} = [draw, very thick, rectangle, minimum size=5.0mm, inner sep=0pt]

		\begin{scope}[xshift=0cm, yshift=-2.5cm]
		\path[->, thick]
		node[format] (a) {$A_1$}
		node[format, below left of=a](b) {$B_1$}
		node[format, below right of=a](c) {$C_1$}
		node[format, right of=a, xshift=-0.4cm] (s) {$S_1$}
		
		(a) edge[blue] (c)
		(a) edge[<->, red] (b)
		(b) edge[blue] (c)		
		
		node[below of=a, yshift=0.0cm, xshift=0.0cm] (la) {$(a) \: s^1$}
		;
		\end{scope}
		
		\begin{scope}[xshift=3cm,yshift=-2.5cm]
		\path[->, thick]
		node[format] (a) {$s^{1}$}
		node[format, below left of=a,,xshift=0cm,yshift=-0.0cm] (b) {$s^{2}$}
		node[format, below right of=a,xshift=0cm,yshift=-0.0cm] (c) {$s^{3}$}
		
		(b) edge[black, bend left=30] (a)
		(a) edge[black, bend left=30] (b)
		(b) edge[black] (c)		
		(c) edge[loop, in=60,out=130,looseness=5] (c)
		
		node[below of=a, yshift=0.0cm, xshift=0.0cm] (la) {$(b)$}
		;
		\end{scope}
		
		\begin{scope}[yshift=-4.7cm, xshift=-0.8cm]
		\path[->, thick]
		node[square] (a1) {$A_{1}$}
		node[square, below of=a1] (b1) {$B_{1}$}
		node[square, below of=b1] (c1) {$C_{1}$}
		node[format, right of=a1, xshift=-0.3cm] (a2) {$A_{12}$}
		node[format, below of=a2, xshift=-0.0cm] (b2) {$B_{12}$}
		node[format, below of=b2, xshift=-0.0cm] (c2) {$C_{12}$}
		node[format, right of=c2, xshift=-0.4cm] (s) {$S_{12}$}
		
		(a2) edge[blue] (b2)
		(b2) edge[blue] (c2)
		
		(c2) edge[blue] (s)
		
		(a1) edge[blue] (a2)
		(c1) edge[blue] (a2)
		(c1) edge[blue] (c2)
		(a2) edge[<->, red, bend left=30] (c2)
		
		node[below of=a1, yshift=-2.0cm, xshift=0.6cm] (la) {$(c) \: \mscriptsize{s^1_{t-1} \rightarrow s^2_t}$}
		;
		\end{scope}
		
		\begin{scope}[yshift=-4.7cm, xshift=2.0cm]
		\path[->, thick]
		node[square] (a1) {$A_{2}$}
		node[square, below of=a1] (b1) {$B_{2}$}
		node[square, below of=b1] (c1) {$C_{2}$}
		node[format, right of=a1, xshift=-0.3cm] (a2) {$A_{23}$}
		node[format, below of=a2] (b2) {$B_{23}$}
		node[format, below of=b2] (c2) {$C_{23}$}
		node[format, right of=c2, xshift=-0.6cm] (s) {$S_{23}$}
		
		
		(a1) edge[blue] (a2)
		(b1) edge[blue] (b2)
		(c1) edge[blue] (c2)
		(c1) edge[blue] (a2)
		
		(a2) edge[blue] (b2)
		(a2) edge[blue, bend left] (c2)
		
		node[below of=a1, yshift=-2.0cm, xshift=0.6cm] (la) {$(d) \:\mscriptsize{ s^2_{t-1} \rightarrow s^3_t}$}
		;
		\end{scope}

		\begin{scope}[yshift=-4.7cm, xshift=4.6cm]
		\path[->, thick]
		node[square] (a1) {$A_{2}$}
		node[square, below of=a1] (b1) {$B_{2}$}
		node[square, below of=b1] (c1) {$C_{2}$}
		node[format, right of=a1, xshift=-0.3cm] (a2) {$A_{21}$}
		node[format, below of=a2] (b2) {$B_{21}$}
		node[format, below of=b2] (c2) {$C_{21}$}
		
		node[format, right of=c2, xshift=-0.6cm] (s) {$S_{21}$}
		
		
		(a1) edge[blue] (a2)
		(b1) edge[blue] (b2)
		(c1) edge[blue] (c2)
		
		(a2) edge[<->,red] (b2)
		(b2) edge[blue] (c2)
		(a2) edge[blue, bend left] (c2)
		
		node[below of=a1, yshift=-2.0cm, xshift=0.6cm] (la) {$(e) \: \mscriptsize{s^2_{t-1} \rightarrow s^1_{t}}$}
		;
		\end{scope}
		
		\begin{scope}[yshift=-4.7cm, xshift=7.0cm, node distance=1.2cm]
		\path[->, thick]
		
		node[format, xshift=0.5cm] (a1) {$A_1$}
		node[format, right of=a1, xshift=-0.0cm] (a2) {$A_{12}$}
		node[format, right of=a2, xshift=-0.0cm] (a3) {$A_{23}$}

		node[format, below of=a1, yshift=0.2cm] (b1) {$B_1$}
		node[format, right of=b1, xshift=-0.0cm] (b2) {$B_{12}$}
		node[format, right of=b2, xshift=-0.0cm] (b3) {$B_{23}$}
		
		node[format, below of=b1, yshift=0.2cm] (c1) {$C_1$}
		node[format, right of=c1, xshift=-0.0cm] (c2) {$C_{12}$}
		node[format, right of=c2, xshift=-0.0cm] (c3) {$C_{23}$}
		
		node[format, below right of=c1, xshift=-0.2cm,yshift=0.3cm] (S1) {$S_1$}
		node[format, below right of=c2, xshift=-0.2cm,yshift=0.3cm] (S2) {$S_2$}
		node[format, below right of=c3, xshift=-0.2cm,yshift=0.3cm] (S3) {$S_3$}
		
		(a1) edge[red, <->] (b1)
		(b1) edge[blue] (c1)
		(a1) edge[blue, bend right=40] (c1)
		
		(a2) edge[<->,red, bend left=40] (c2)
		(b2) edge[blue] (c2)
		(a2) edge[blue] (b2)
		
		(a3) edge[blue] (b3)
		(a3) edge[blue, bend left=40](c3)

		(a1) edge[blue] (a2)
		(a2) edge[blue] (a3)
		(b2) edge[blue] (b3)
		(c1) edge[blue] (c2)
		(c2) edge[blue] (c3)
		
		(c1) edge[blue] (a2)
		(c2) edge[blue, bend left=5] (a3)
		
		(c2) edge[blue] (S2) 
		node[below of=a1, yshift=-2.3cm, xshift=1.6cm] (la) {$(f)$}
		;
		\end{scope}

		\end{tikzpicture}
	\end{center}
	\caption {A hidden variable PDSEM.
		(a) Causal structure of the initial state $S^1$.
		(b) The state transition diagram.
		(c),(d),(e) Latent projected causal diagrams representing possible transitions and subsequent states.
		(f) A snapshot of a possible PDSEM trajectory represented as an unrolled ADMG}
	\label{fig:admg-pdsem-example}
\end{figure}
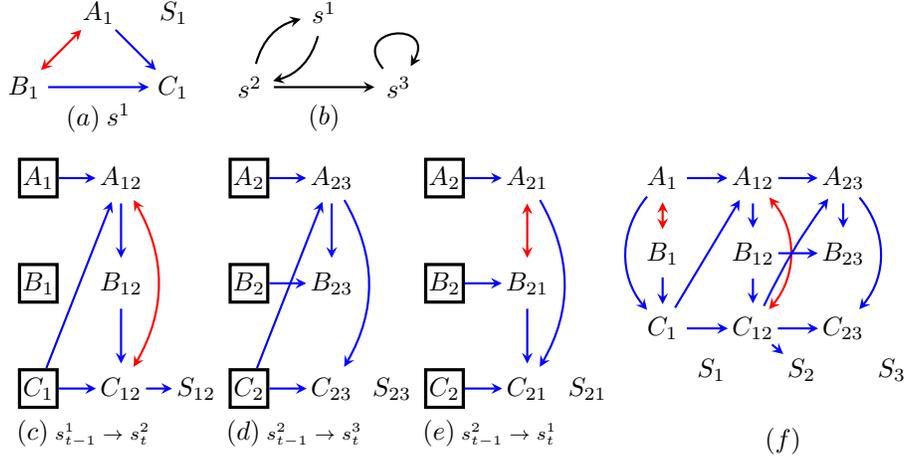

\section{Proofs}\label{proofs}
\begin{lema}{\ref{eqn:hidden-var-causal-dbn}}
	Under Assumption \ref{assump:fixed-observed}, $p({\bf Y}({\bf a}))$ is identified from a hidden variable causal DBN model represented by latent projections ${\cal G}_1$ on ${\bf V}_1$ and ${\cal G}_{t+1,t}$ on ${\bf V}_{t+1}$ given ${\bf V}_t$ if and only if every bidirected connected component in ${{\cal G}_1}_{{\bf Y}^*_1}$ (the induced subgraph of ${\cal G}_1$) is intrinsic in ${\cal G}_1$, and every bidirected component in ${{\cal G}_{t+1,t}}_{{\bf Y}^*_{i}}$ (the induced subgraph of ${\cal G}_{t+1,t}$) is intrinsic in ${\cal G}_{t+1,t}$, where ${\bf Y}^*_1$ is the set of ancestors of ${\bf Y} \cap {\bf V}_1$ not through ${\bf A} \cap {\bf V}_1$ in ${\cal G}_1$, and for every $i \in 2,\ldots, T$, ${\bf Y}^*_{i}$ is the set of ancestors of ${\bf Y} \cap {\bf V}_{i}$ not through ${\bf A} \cap {\bf V}_{i}$ in ${\cal G}_{t+1,t}$.  Moreover, if 
	$p({\bf Y}({\bf a}))$ is identified, we have
	\vspace*{-1mm}
	{\small
		\begin{align*}
		\Bigg( \sum_{{\bf Y}^*_1 \setminus (({\bf Y} \cup {\bf A}) \cap {\bf V}_1)} 
		\prod_{{\bf D} \in {\cal D}({{\cal G}_1}_{{\bf Y}^*_1})} q^1_{\bf D}({\bf D} | \pa_{\cal G}({\bf D}) \setminus {\bf D}) \vert_{{\bf A} = {\bf a}}
		\Bigg) \times\\
		\prod_{i=2}^T
		\Bigg( \sum_{{\bf Y}^*_i \setminus (({\bf Y} \cup {\bf A}) \cap {\bf V}_i)} 
		\prod_{{\bf D} \in {\cal D}({{\cal G}_{t+1,t}}_{{\bf Y}^*_i})} q^{t+1,t}_{\bf D}({\bf D} | \pa_{\cal G}({\bf D}) \setminus {\bf D}) \vert_{{\bf A} = {\bf a}},
		\Bigg)
		\end{align*}
	}%
	where $q^1_{\bf D}$ and $q^{t+1,t}_{\bf D}$ are kernels corresponding to intrinsic sets representing elements of ${\cal D}({{\cal G}_1}_{{\bf Y}^*_1})$ and ${\cal D}({{\cal G}_{t+1,t}}_{{\bf Y}^*_1})$ in the nested Markov factorizations of ${\cal G}_1$ and ${\cal G}_{t+1,t}$, respectively.
	
\end{lema}
\begin{prf}
	We want to obtain $p({\bf Y}({\bf a}))$ from the observed joint $p({\bf V}_{1:T})$. Using identification result \ref{eqn:id} on the unrolled ADMG gives 	
	$\sum_{Y^* \setminus Y} p({\bf Y}^*(a)) = \sum_{{\bf Y}* \setminus {\bf Y}} \prod_{D \in \mathcal{D}(\G_{\text{unrolled} {\bf Y}^*})} p({\bf D}(\pa({\bf D})\setminus {\bf D})) |_{{\bf A}={\bf a}}$. Assumption \ref{assump:fixed-observed} ensures that no district $\bf D$ spans time points, and parents $\pa(\bf D)$ at time $t+1$ lie either at $t$ or $t+1$.
	This allows us to write $\sum_{{\bf Y}^* \setminus {\bf Y}} p({\bf Y}^*({\bf a})) = \sum_{{\bf Y}^* \setminus {\bf Y}} \prod_{{\bf D} \in \mathcal{D}(\G_{1Y^*})} p({\bf D}(\pa({\bf D})\setminus {\bf D})) |_{{\bf A}={\bf a}} \times \prod_{t=1}^{T-1} \prod_{{\bf D} \in \mathcal{D} (\G_{t+1,t Y^*})} p({\bf D}(\pa({\bf D})\setminus {\bf D})) |_{{\bf A}={\bf a}}$. Applying the identification results in \cite{richardson12nested} to the prior network ADMG $\G_1$ and extensions of these results in \cite{sherman2018identification} to the transition network CADMGs $\G_{t+1,t}$, these counterfactual conditionals can be replaced by given modified nested factorizations, provided every appropriate bidirected connected set in the prior or transition graph is intrinsic in that graph.
\end{prf}

\begin{lema}{\ref{eqn:pdsem-g}}
	Given a fully observed PDSEM, each factor of the distribution $p_{\infty}({\bf Y}({\bf a}))$ is identified from $p_{\infty}({\bf V})$ as:
	{\small
		\begin{align*}
		\notag
		p_1({\bf Y}_1({\bf a}_1)) &\equiv \prod_{V \in {\bf Y}_1 \setminus {\bf A}_1} p_1(V | \pa_{{\cal G}_1}(V)) \Big\vert_{{\bf A}_1 = {\bf a}_1} \\
		p_{ij}({\bf Y}_{ij}({\bf a}_{j}) | {\bf Y}_i({\bf a}_i)) &\equiv \prod_{V \in {\bf Y}_{ij} \setminus {\bf A}_{j}} p_{ij}(V | \pa_{{\cal G}_{ij}}(V)) \Big\vert_{\substack{{\bf A}_i = {\bf a}_i, \\{\bf A}_j = {\bf a}_j}}
		\end{align*}
	}
\end{lema}
\begin{prf}
	This follows from the factorization of $p_{\infty}({\bf V}({\bf a}))$ into elements of the form $p_1({\bf Y}_1({\bf a}_1))$, and $ p_{ij}({\bf Y}_{j}({\bf a}_{j}) | {\bf Y}_i(\bf a_i)) $, the fact that
	${\cal G}_1, \{ {\cal G}_{ij} : (i,j) \in {\cal T} \}$ define causal models under standard structural equation semantics, and equation \ref{eqn:g} .
\end{prf}

\begin{lema} {\ref{eqn:pdsem-admg-id}}
Under Assumptions \ref{assump:fixed-observed}, \ref{assump:state-vars} and \ref{assump:transition-observed}, given a latent variable PDSEM represented by ${\cal G}_1$ and $\{ {\cal G}_{ij} : (i,j) \in {\cal T} \}$,
$p_{\infty}({\bf Y}({\bf a}))$ is identified from $p_{\infty}({\bf V})$ if and only if every bidirected component in ${\cal G}_{1{\bf Y}_1}$ is intrinsic in ${\cal G}_1$, and
every bidirected component in ${\cal G}_{ij{\bf Y}_j}$ is intrinsic in ${\cal G}_{ij}$ for every $i$ and $j$.  Moreover, if $p_{\infty}({\bf Y}({\bf a}))$ is identified, it is equal to
{\small
\begin{gather}
p_1({\bf Y}_1({\bf a}_1)) \prod_{t=1}^{\infty} \left( \prod_{(i,j) \in {\cal T}} \left( p_{ij}({\bf Y}_{ij}({\bf a}_{j}) | {\bf Y}_i({\bf a}_i)) \right)^{\mathbb{I}(s^i_{t-1},s^j_{t})} \right)
\!\!
1^{\mathbb{I}(s_{t-1}^*)}
\label{eqn:admg-pdsem-top}
\shortintertext{where}
p_1({\bf Y}_1({\bf a}_1)) = \prod_{{\bf D} \in {\cal D}({\cal G}_{1{\bf Y}^*_1})} q^1_{\bf D}({\bf D} | \pa^s_{{\cal G}_1}({\bf D})) \Big\vert_{{\bf A}_1 = {\bf a}_1},
\label{eqn:admg-pdsem-prior}
\end{gather}}%
\vspace*{-0.6mm}%
where each kernel $q^1_{\bf D}({\bf D} | \pa^s_{{\cal G}_1}({\bf D}))$ is in the nested Markov factorization of $p_1({\bf V}_1)$ with respect to ${\cal G}_1$, and
{\small
\begin{align}
p_{ij}({\bf Y}_{ij}({\bf a}_{j}) | {\bf Y}_i({\bf a}_i)) = \quad \prod_{\mathclap{{\bf D} \in {\cal D}({\cal G}_{{\bf V}_{ij} \setminus {\bf A}_{ij}})}} \quad q^{ij}_{\bf D}({\bf D} | \pa^s_{{\cal G}_{ij}}({\bf D})) \Big\vert_{\substack{{\bf A}_i={\bf a}_i, \\ {\bf A}_j={\bf a}_j}}
\label{eqn:admg-pdsem-transition}
\end{align}
}\vspace*{-0.5mm}%
where each kernel $q^{ij}_{\bf D}({\bf D} | \pa^s_{{\cal G}_{ij}}({\bf D}))$ is in the nested Markov factorization of $p_{ij}({\bf V}_{ij} | {\bf V}_i)$ with respect to 
${\cal G}_{ij}$.
\end{lema}

\begin{prf}
Assumption \ref{assump:transition-observed} implies all state transitions are known, and thus allows us to proceed by induction on any sequence of state transitions with positive probability after $t$ steps.
	
Unrolling the prior network, and appropriate transition networks for such a sequence yields an ADMG representing the observed data distribution had that transition taken place, with Assumption \ref{assump:fixed-observed} implying that districts in this ADMG do not span multiple time steps.  This immediately implies the conclusion by the same argument used in the proof of Lemma~\ref{eqn:hidden-var-causal-dbn}.

In fact, this argument works for any transition sequence of any size.	
\end{prf}

\section{The Septoplasty Surgical Procedure, and its PDSEM Model}\label{septoplasty}

Septoplasty is a surgical procedure performed on the nasal cartilage, called the septum, to relieve nasal obstruction \citep{tajudeen2017thirty}. A deviated or deformed septum is the most common cause of such an obstruction. Apart from nasal obstruction, a significantly deviated nasal septum has also been implicated in epistaxis, sinusitis, obstructive sleep apnea, and headaches which can act as diagnosis factors. The procedure involves cartilage resection, modification or a graft. The outcome of septoplasty is typically a score/index constructed from a questionnaire investigating quality of life measures and perceived nasal obstruction levels, like Nasal Obstruction Septoplasty Effectiveness (NOSE) and the Fairley Nasal Questionnaire (FNQ) \citep{fettman2009surgical}.

For instructional and evaluation purposes, surgeries are often divided into discrete steps or "stages", each with its own intermediate goal \citep{ahmidi2015automated}. 
Our data from the septoplasty procedure was manually annotated by clinical experts and divided into the following states:
\begin{itemize}
	\item $s_1$: opening of the septum,
	\item $s_2$: raising septal flaps,
	\item $s_3$: removal of deviated septal cartilage and bone,
	\item $s_4$: reconstruction,
	\item $s_5$: closing of the incision,
	\item $s_6$: activity not otherwise included in the above 5 phases,
	\item $s_{\textrm{end}}$: end of surgery state (which contains no variables).
\end{itemize}   

The variables in our data are the following: $\mathbf{V} = \{$ K: knife, G: gorney scissors, C$_1$: cottle, D$_1$: short needle driver, D$_2$: long needle driver, O: other tools, C$_2$: suction cannula, M: main surgeon exists, S: suction exists, A$_1$: main surgeon is an attending, A$_2$: suction done by attending, T: duration of that phase is greater than 10 seconds $\}$

\begin{itemize}
	\item $\mathbf{V}_{s_1} = \{\textrm{K, O, C}_2 \textrm{, M, S, A}_1\textrm{, A}_2\textrm{, T}\}$,
	\item $\mathbf{V}_{s_2} = \{\textrm{K, C}_1\textrm{, O, C}_2 \textrm{, M, S, A}_1 \textrm{, A}_2 \textrm{, T}\}$,
	\item $\mathbf{V}_{s_3} = \{\textrm{K, C}_1 \textrm{, D}_1 \textrm{, D}_2 \textrm{, O, C}_2 \textrm{, M, S, A}_1\textrm{, A}_2\textrm{, T}\}$,
	\item $\mathbf{V}_{s_4} = \{\textrm{K, C}_1\textrm{, G, O, C}_2\textrm{, M, S, A}_1\textrm{, A}_2\textrm{, T}\}$,
	\item $\mathbf{V}_{s_5} = \{\textrm{D}_1\textrm{, D}_2\textrm{, O, C}_2\textrm{, M, S, A}_1\textrm{, A}_2\textrm{, T}\}$,
	\item $\mathbf{V}_{s_6} = \{\textrm{K, C}_1\textrm{, O, C}_2\textrm{, M, S, A}_1\textrm{, A}_2\textrm{, T}\}$,
\end{itemize}

To determine the allowed state transitions , we retained observed data state transitions where at least 5 such transitions occurred. The permitted state transitions $s_i \to s_j$ are summarized in Figure \ref{fig:transition-diagram-data} in the main paper -- note that transitions other than those depicted have  probability $p(s_j \mid s_i, \mathbf{v}_{s_i}) = 0 $ for all $\mathbf{v}_{s_i}$. To determine the state transition distributions $p(s_j \mid s_i, \mathbf{v}_{s_i})$, we restricted the set $\mathbf{v}_{s_i}$ for all $i$ to be $\{A_1, M, A_2, S\}$ to increase tractability of estimation, and estimated this discrete conditional distribution via a conditional probability table. The prior distribution on the initial state was set to $p(s_1) = 1$. 

State DAGs  were determined based on clinician recommendation and have been reproduced in Figure \ref{fig:septo-graphs} for reference. These immediately lead to prior variable distributions $p(\mathbf{V}_{s_i})$ for each state $s_i$.

Transition graphs from $s_i \to s_j$ are constructed using a simple rule: the for any variable $v$ in any state $s_j$, the parents $\pa(v_{s_j})$ consists of the variable with the same name in the previous state $s_i$ if it exists, and all parents in the state DAG for $s_j$  point indicated by state DAGs. For example, in the transition $s_1 \to s_1$ moving from time step $t-1 \to t$, variable $K$ at time step $t$ has parents $A_1$ at $t$, as given in Figure \ref{fig:septo-graphs}(a), as well as $K$ from time step $t-1$. However, in the transition $s_1 \to s_2$ moving from time step $t-1$ to time step $t$, variable $C_1$ has parent $A_1$ in time step $t$, as given in Figure \ref{fig:septo-graphs}(b), but no parents from the previous time step $t-1$ since $C_1$ does not exist in $s_1$. Based on this rule, probability distributions $p(v_{s_j} \mid \pa(v_{s_j}))$ are estimated using conditional probability tables.


\begin{figure}  %
	\begin{center}
		\begin{tikzpicture}[>=stealth, node distance=1.2cm]
		\tikzstyle{format} = [draw=none, very thick, circle, minimum size=7.0mm,
		inner sep=0pt]
		
		\begin{scope}[xshift=-10cm,yshift=-2.5cm]
		\path[->, very thick]
		
		node (m) {$M$}
		node[below of=m] (s) {$S$}
		node[format, right of=m] (a1) {$A_1$}
		node[format, right of=s] (a2) {$A_2$}
		node[format, right of=a1,yshift=-0.5cm] (k) {$K$}
		node[format, above of=k] (o) {$O$}
		node[format, below of=k] (c2) {$C_2$}
		node[format, right of=k,yshift=-0.5cm,xshift=-0cm] (t) {$T$}

		(m) edge[blue] (s)
		(a1) edge[blue] (a2)
		(m) edge[blue] (a1)
		(s) edge[blue] (a2)
		(a1) edge[blue] (o)
		(a1) edge[blue] (k)
		(a2) edge[blue] (c2)
		(o) edge[blue, bend left=20] (t)
		(k) edge[blue] (t) 
		(c2) edge[blue, bend right=20] (t)
		(a1) edge[blue, bend left=30] (t)
		(a2) edge[blue, bend right=0] (t)
		node[below of=m, yshift=-1.5cm, xshift=1.5cm] (l) {$(a)$}
		;
		\end{scope}
	
		\begin{scope}[xshift=-5cm,yshift=-2.5cm]
		\path[->, very thick]
		
		node (m) {$M$}
		node[below of=m] (s) {$S$}
		node[format, right of=m] (a1) {$A_1$}
		node[format, right of=s] (a2) {$A_2$}
		node[format, right of=a1] (c1) {$C_1$}
		node[format, right of=a2] (k) {$K$}		
		node[format, right of=a1,yshift=1cm] (o) {$O$}
		node[format, right of=a2,yshift=-1cm] (c2) {$C_2$}
		node[format, right of=c1,yshift=-0.5cm,xshift=-0cm] (t) {$T$}

		(m) edge[blue] (s)
		(a1) edge[blue] (a2)
		(m) edge[blue] (a1)
		(s) edge[blue] (a2)
		(a1) edge[blue] (o)
		(a1) edge[blue] (c1)
		(a1) edge[blue] (k)
		(a2) edge[blue] (c2)
		(o) edge[blue, bend left=20] (t)
		(c1) edge[blue] (t)
		(k) edge[blue] (t)
		(c2) edge[blue, bend right=20] (t)
		(a1) edge[blue, bend left=40] (t)
		(a2) edge[blue, bend right=40] (t)
		node[below of=m, yshift=-1.5cm, xshift=1.5cm] (l) {$(b)$}
		;
		\end{scope}
		
		\begin{scope}[xshift=0cm,yshift=-2.5cm]
		\path[->, very thick]
		
		node (m) {$M$}
		node[below of=m] (s) {$S$}
		node[format, right of=m] (a1) {$A_1$}
		node[format, right of=s] (a2) {$A_2$}
		node[format, right of=a1] (c1) {$C_1$}
		node[format, right of=a2, yshift=0.55cm] (g) {$G$}
		node[format, right of=a2, yshift=-0.15cm] (k) {$K$}		
		node[format, right of=a1,yshift=1cm] (o) {$O$}
		node[format, right of=a2,yshift=-1cm] (c2) {$C_2$}
		node[format, right of=c1,yshift=-0.5cm,xshift=-0cm] (t) {$T$}

		(m) edge[blue] (s)
		(a1) edge[blue] (a2)
		(m) edge[blue] (a1)
		(s) edge[blue] (a2)
		(a1) edge[blue] (o)
		(a1) edge[blue] (c1)
		(a1) edge[blue] (g)
		(a1) edge[blue] (k)
		(a2) edge[blue] (c2)
		(o) edge[blue, bend left=20] (t)
		(c1) edge[blue] (t)
		(k) edge[blue] (t)
		(g) edge[blue] (t)
		(c2) edge[blue, bend right=20] (t)
		(a1) edge[blue, bend left=40] (t)
		(a2) edge[blue, bend right=50] (t)
		node[below of=m, yshift=-1.5cm, xshift=1.5cm] (l) {$(c)$}
		;
		\end{scope}
		
		\begin{scope}[xshift=-10cm,yshift=-7.5cm]
		\path[->, very thick]
		
		node (m) {$M$}
		node[below of=m] (s) {$S$}
		node[format, right of=m] (a1) {$A_1$}
		node[format, right of=s] (a2) {$A_2$}
		node[format, right of=a1] (c1) {$C_1$}
		node[format, right of=a2, yshift=0.55cm] (d1) {$D_1$}
		node[format, right of=a2, yshift=-0.15cm] (k) {$K$}		
		node[format, right of=a1,yshift=1.5cm] (o) {$O$}
		node[format, right of=a1,yshift=0.75cm] (d2) {$D_2$}
		node[format, right of=a2,yshift=-1cm] (c2) {$C_2$}
		node[format, right of=c1,yshift=-0.5cm,xshift=-0cm] (t) {$T$}

		(m) edge[blue] (s)
		(a1) edge[blue] (a2)
		(m) edge[blue] (a1)
		(s) edge[blue] (a2)
		
		(a1) edge[blue] (o)
		(a1) edge[blue] (c1)
		(a1) edge[blue] (d1)
		(a1) edge[blue] (k)
		(a1) edge[blue, bend left=10] (d2)
		
		(a2) edge[blue] (c2)
		
		(o) edge[blue, bend left=20] (t)
		(c1) edge[blue] (t)
		(k) edge[blue] (t)
		(d1) edge[blue] (t)
		(d2) edge[blue, bend left=20] (t)
		(c2) edge[blue, bend right=20] (t)
		
		(a1) edge[blue, bend left=40] (t)
		(a2) edge[blue, bend right=50] (t)

		node[below of=m, yshift=-1.5cm, xshift=1.5cm] (l) {$(d)$}
		;
		\end{scope}

		\begin{scope}[xshift=-5cm,yshift=-7.5cm]
		\path[->, very thick]
		
		node (m) {$M$}
		node[below of=m] (s) {$S$}
		node[format, right of=m] (a1) {$A_1$}
		node[format, right of=s] (a2) {$A_2$}
		node[format, right of=a1] (d1) {$D_1$}
		node[format, right of=a2] (d2) {$D_2$}		
		node[format, right of=a1,yshift=1cm] (o) {$O$}
		node[format, right of=a2,yshift=-1cm] (c2) {$C_2$}
		node[format, right of=d1,yshift=-0.5cm,xshift=-0cm] (t) {$T$}

		(m) edge[blue] (s)
		(a1) edge[blue] (a2)
		(m) edge[blue] (a1)
		(s) edge[blue] (a2)
		(a1) edge[blue] (o)
		(a1) edge[blue] (d1)
		(a1) edge[blue] (d2)
		(a2) edge[blue] (c2)
		(o) edge[blue, bend left=20] (t)
		(d1) edge[blue] (t)
		(d2) edge[blue] (t)
		(c2) edge[blue, bend right=20] (t)
		(a1) edge[blue, bend left=40] (t)
		(a2) edge[blue, bend right=40] (t)
		node[below of=m, yshift=-1.5cm, xshift=1.5cm] (l) {$(e)$}
		;
		\end{scope}
		
		\begin{scope}[xshift=0cm,yshift=-7.5cm]
		\path[->, very thick]
		
		node (m) {$M$}
		node[below of=m] (s) {$S$}
		node[format, right of=m] (a1) {$A_1$}
		node[format, right of=s] (a2) {$A_2$}
		node[format, right of=a1] (c1) {$C_1$}
		node[format, right of=a2] (k) {$K$}		
		node[format, right of=a1,yshift=1cm] (o) {$O$}
		node[format, right of=a2,yshift=-1cm] (c2) {$C_2$}
		node[format, right of=c1,yshift=-0.5cm,xshift=-0cm] (t) {$T$}

		(m) edge[blue] (s)
		(a1) edge[blue] (a2)
		(m) edge[blue] (a1)
		(s) edge[blue] (a2)
		(a1) edge[blue] (o)
		(a1) edge[blue] (c1)
		(a1) edge[blue] (k)
		(a2) edge[blue] (c2)
		(o) edge[blue, bend left=20] (t)
		(c1) edge[blue] (t)
		(k) edge[blue] (t)
		(c2) edge[blue, bend right=20] (t)
		(a1) edge[blue, bend left=40] (t)
		(a2) edge[blue, bend right=40] (t)
		node[below of=m, yshift=-1.5cm, xshift=1.5cm] (l) {$(f)$}
		;
		\end{scope}

		\end{tikzpicture}
	\end{center}
	\caption{(a)-(f) State DAGs corresponding to states $s^1$: Opening of the septum, $s^2$: Raising septal flaps, $s^3$: Removal of deviated septal cartilage and bone, $s^4$: Reconstruction, $s^5$: Closing of incision and $s^6$: Activity not otherwise included in the above phases.}
	\label{fig:septo-graphs}
\end{figure}
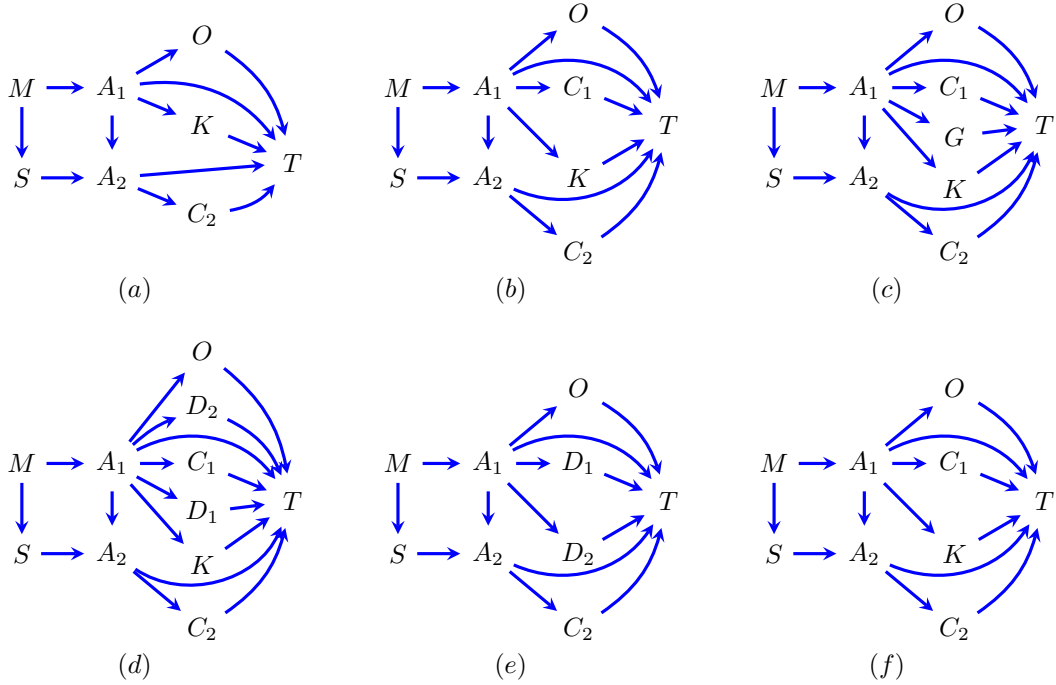

Goodness of fit of our model with respect to the original data distribution is shown in Figure \ref{fig:real-vs-model}. Trajectories simulated by our model are able to capture the distribution of surgery duration originally seen in the data, quite well.
 
\begin{figure}[h]
	\begin{center}
		\centerline{\includegraphics[width=0.6\columnwidth]{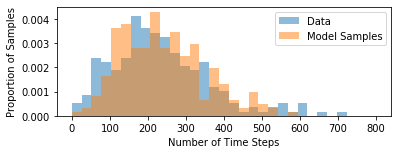}}
		\caption{Histograms of observed surgery (blue) versus simulated surgeries from the estimated model (orange).}
		\label{fig:real-vs-model}
	\end{center}
	\vskip -0.2in
\end{figure}

{\bf Implications for Statistical Inference For Latent Variable PDSEMs:}
If a PDSEM is fully observed, causal inference may be performed by obtaining maximum likelihood estimates $\hat{\eta}$ of all parameters, and evaluating the g-formula functionals using Monte-Carlo sampling using learned distributions of the form $p(V | \pa_{\cal G}(V); \eta_V)$.  This method is computationally efficient as long as the initial DAG and transition CDAGs in a PDSEM are sufficiently sparse.  Indeed, our data application was based on this approach.

However, an analogous approach is not straightforward for nested Markov parameterizations of the marginal PDSEM representing a PDSEM with hidden variables. In our simulations, we use a specific generative model for our continuous variables, i.e, the linear Gaussian Structural Equation model. Another choice based on work in \citep{evans14markovian} is the M\"{o}bius parameterization for binary variables. However, this is ill-suited for drawing samples. Instead, existing approaches to sampling from a nested Markov discrete likelihood involve first converting the likelihood expressed in terms of the M\"{o}bius parameters to one expressed as a the joint distribution $p({\bf V})$ (from which it is easy to generate samples for a discrete sample space of ${\bf V}$).  Importantly, such a conversion leads to an intractable object that requires storage and running time exponential in $|{\bf V}|$.  This holds \emph{even if} the underlying model dimension of the nested Markov model is small.  The situation is radically different from that of DAG models, where a small model dimension directly leads to a computationally efficient sampling scheme. For settings beyond Gaussian and discrete data, statistical inference strategies are significantly more complicated and have been discussed in \cite{rozi2020semiparam}.

While there exist promising approaches, based on the nested Markov generalization of the variable elimination algorithm \citep{shpitser11eid}, in general the problem remains open.

\section{Computation Details}

The septoplasty data application presented in Section \ref{experiments} was computed on a Lenovo X1 Carbon with an Intel i7 1.8 GHz processor and 16 GB of RAM. Computation for each scenario (generating from the model without interventions, attending performing the whole surgery, and trainee performing the whole surgery) took between 1.5 to 2 hours each.

%

\end{document}